\documentclass[reprint,amsmath,amssymb,aps,superscriptaddress]{revtex4-2}
\usepackage[colorlinks,bookmarks=true,citecolor=blue,linkcolor=blue,urlcolor=blue]{hyperref}
\usepackage{dcolumn,graphicx,amsfonts,amsthm,bm,color,appendix,float}

\newcommand{\mC}{\mathcal{C}}

\newcommand{\mB}{\mathcal{B}}
\newcommand{\mF}{\mathcal{F}}
\newcommand{\tmH}{\tilde{\mathcal H}}
\newcommand{\tbm}[1]{\tilde{\bm #1}}
\newcommand{\mr}{moir\'e~}

\begin{document}
\title{Origin of Model Fractional Chern Insulators in All Topological Ideal Flatbands: Explicit Color-entangled Wavefunction and Exact Density Algebra}

\author{Jie Wang}
\email{jiewang.phy@gmail.com}
\affiliation{Center for Mathematical Sciences and Applications, Harvard University, Cambridge, MA 02138, USA}
\affiliation{Department of Physics, Harvard University, Cambridge, MA 02138, USA}
\affiliation{Center for Computational Quantum Physics, Flatiron Institute, 162 5th Avenue, New York, NY 10010, USA}

\author{Semyon Klevtsov}
\affiliation{IRMA, Universit\'e de Strasbourg, UMR 7501, 7 rue Ren\'e Descartes, 67084 Strasbourg, France}

\author{Zhao Liu}
\email{zhaol@zju.edu.cn}
\affiliation{Zhejiang Institute of Modern Physics, Zhejiang University, Hangzhou 310027, China}

\begin{abstract}
It is commonly believed that nonuniform Berry curvature destroys the Girvin-MacDonald-Platzman algebra and as a consequence destabilizes fractional Chern insulators. In this work we disprove this common lore by presenting a theory for all topological ideal flatbands with nonzero Chern number $\mC$. The smooth single-particle Bloch wavefunction is proved to admit an exact color-entangled form as a superposition of $\mC$ lowest Landau level type wavefunctions distinguished by boundary conditions. Including repulsive interactions, Abelian and non-Abelian model fractional Chern insulators of Halperin type are stabilized as exact zero-energy ground states no matter how nonuniform Berry curvature is, as long as the quantum geometry is ideal and the repulsion is short-ranged. The key reason behind is the existence of an emergent Hilbert space in which Berry curvature can be exactly flattened by adjusting wavefunction's normalization. In such space, the flatband-projected density operator obeys a closed Girvin-MacDonald-Platzman type algebra, making exact mapping to $\mC-$layered Landau levels possible. In the end we discuss applications of the theory to \mr flatband systems with a particular focus on the fractionalized phase and spontaneous symmetry breaking phase recently observed in graphene based twisted materials.
\end{abstract}

\maketitle


\section{Introduction}
Flatbands are ideal platforms for realizing exotic quantum phases of matter. Of particular interests are topological flatbands where fractionalized topological phases are possible~\cite{QH_RMP,RevModPhys.71.S298,PhysRevX.1.021014,PARAMESWARAN2013816,zhao_review,Neupertreview_2015,zhao_chapter}. During recent years, there has been numerous progress in creating and engineering flatband systems such as by stacking and twisting~\cite{Andrei:2020aa,Bistritzer12233,Cao:2018aa,Cao:2018ab}. Understanding the stability of fractionalized phases with respect to tuning parameters such as energy dispersion, wavefunction geometry, interaction range is crucial for material engineering, experimental realization and theoretical characterization of these exotic phases. Along this path, exact statements are particularly important. In this work, we present an exact condition for the stability of fractional Chern insulators (FCIs), {\it i.e.} the lattice version of fractional quantum Hall (FQH) states~\cite{PhysRevX.1.021014,PARAMESWARAN2013816,zhao_review,Neupertreview_2015,zhao_chapter}. Our results also provide a framework to systematically explore the interplay between wavefunction's geometry and interactions, and are useful in guiding material engineering toward realizing exotic quantum phases of matter.

Landau levels (LLs) are the simplest topological Chern flatbands, with unit Chern number $\mC=1$ and exactly zero bandwidth. They are well known for their extreme uniformness that is crucial for the stability of FQH states: LLs are covariant under any smooth area preserving deformations~\cite{Winfinty_Cappelli,Winfinty_Karabali} which is encoded in the LL projected density algebra initially noticed by Girvin, MacDonald, and Platzman (GMP) for lowest Landau level (LLL) states~\cite{gmpl,gmpb}:
\begin{equation}
    ~[\hat\rho_{\bm q_1}, \hat\rho_{\bm q_2}] = \left(e^{q^*_1q_2 l_B^2}- e^{q_1q^*_2 l_B^2}\right)\hat\rho_{\bm q_1+\bm q_2},\label{projectedGMP}
\end{equation}
where $\hat\rho_{\bm q}$ is the LLL projected density operator, $q$ is the complex coordinate of momentum which is $q = (q_x + iq_y)/\sqrt2$ in an isotropic LL, and $l_B$ is the magnetic length. The density operator deforms the shape of LLs while preserving their LL index~\cite{gmpl,gmpb,Winfinty_Cappelli,Winfinty_Karabali}. The GMP algebra is important for FQH physics in many aspects including the conformal field theory mapping~\cite{MoreReadState,hierarchyreview}, constructing exact pseudopotential projectors that protect the stability of FQH states~\cite{Haldane_hierarchy,TrugmanKivelson85} and others. The GMP algebra was initially derived based on the holomorphicity of LLL wavefunctions~\cite{girvin_holomorphic} which arises from the non-commutativity of guiding centers and in fact applies to any LL~\cite{Haldanegeometry}.

Topological flatbands realized in lattice systems are more complicated than LLs. They can carry arbitrary integer Chern number, and typically do not exhibit uniform Berry curvature. Due to this complication, neither the holomorphicity of wavefunction nor the exact projected density algebra exist, thereby Eq.~(\ref{projectedGMP}) becomes approximate. As a consequence FCIs are no longer expected to be exact and their stability is supposed to be reduced~\cite{Jackson:2015aa}. There have been important attempts towards restoring the GMP algebra in flatband systems. For instance, Ref.~\cite{PhysRevB.90.165139} highlighted the importance of the Fubini-Study metric $g^{ab}_{\bm k}$ which is the intrinsic wavefunction distance measure. Given constant Berry curvature $\Omega_{\bm k} = \Omega$ and the so-called trace condition $\text{Tr} g_{\bm k} = \Omega_{\bm k}$, Ref.~\cite{PhysRevB.90.165139} derived a GMP algebra for all topological flatbands of $\mC\neq0$. However, the assumption of constant Berry curvature oversimplified the flatband problem.

In this work, we relax the above assumption by focusing on ideal flatbands satisfying the following conditions~\cite{JieWang_exactlldescription}:
\begin{equation}
    g^{ab}_{\bm k} = \frac{1}{2}\omega^{ab}\Omega_{\bm k},\quad\Omega_{\bm k} > 0\quad \text{for}\quad \forall\bm k.\label{defidealcond}
\end{equation}
The $\omega_{ab}$ is a constant uni-modular matrix. The Berry curvature is assumed to be positive definite but not necessarily uniform. Recently, it is known that Eq.~(\ref{defidealcond}) is necessarily~\cite{Martin_PositionMomentumDuality} and sufficiently~\cite{kahlerband1,kahlerband2,kahlerband3} equivalent to the momentum-space holomorphicity, which is a key geometric property of the momentum-space (= boundary-condition space). For all $\mC=1$ ideal flatbands such as those realized in the chiral model of twisted bilayer graphene~\cite{Grisha_TBG,Grisha_TBG2} and others~\cite{Kapit_Mueller,Emil_constantBerry,Liang_DiracNonuniformB}, it has been shown that momentum-space holomorphicity highly constrains the wavefunction to admit a universal form descending from the LLL wavefunction~\cite{JieWang_exactlldescription}. Such simplicity enables exact construction of many-body model wavefunctions~\cite{Grisha_TBG2} as well as exact projective pseudopotential Hamiltonians~\cite{JieWang_exactlldescription}, hence the ideal condition plays an importance role in experimental realization and identification of FCIs~\cite{FCI_TBG_exp,Dan_parker21}. Recently the ideal condition Eq.~(\ref{defidealcond}) has been generalized to a larger family called ``vortexability''~\cite{LedwithVishwanathParker22} by allowing nonlinear real-space embedding~\footnote{To clarify different terminologies: ``Kahler band''~\cite{kahlerband1,kahlerband2,kahlerband3} refers to a system satisfying Eq.~(\ref{defidealcond}) locally for every $\bm k$ where $\omega^{ab}_{\bm k}$ can be momentum dependent; ``ideal band''~\cite{JieWang_exactlldescription} requires $\omega^{ab}$ to be $\bm k-$independent; ``flat Kahler band'' further demands constant Berry curvature; ``vortexable band''~\cite{LedwithVishwanathParker22} allows non-linear embeding in the real-space.}.

While much progress are made for $\mC=1$, less is known for generic cases of $\mC>1$. Recently, solvable models based on twisted multilayer graphene sheets were proposed which realize ideal flatbands of arbitrary Chern number~\cite{Jie_hierarchyidealband,Eslam_highC_idealband}. Remarkably, exact Halperin type FCIs were found by numerical diagonalization~\cite{Jie_hierarchyidealband}. Motivated by the $\mC=1$ case, it was conjectured that the momentum-space holomorphicity is also the fundamental reason for their emergence~\cite{Jie_hierarchyidealband}, however, the nature of the Bloch wavefunction and why FCIs are stable are still far from thorough understanding.

In this work, we prove that the ideal quantum geometric condition, without assuming the flatness of $\Omega_{\bm k}$ or $g_{\bm k}$ themselves, is sufficient to guarantee exact GMP algebra and model FCI states occurring as exact zero-energy ground states of proper short-ranged interactions in all $\mC\geq 1$ ideal flatbands. We achieve these results by pointing out the importance of an emergent Hilbert space in which the wavefunctions' normalization factors are adjusted to flatten Berry curvature in an exact manner. Importantly, this leads to a simple algebra for the projected coordinates identical to the guiding center algebra in LLs, and enables an exact derivation for the single-particle wavefunction in ideal flatbands. The general form Bloch wavefunction is found to be a nonlinear superposition of LLL wavefunctions of $\mC$ distinct boundary conditions, generalizing the previously proposed color-entangled wavefunctions with uniform Berry curvature~\cite{Maissam_Qi_PRX12,Yangle_Modelwf,Yangle_haldanestatistics} to the case of fluctuating Berry curvature. The density algebra is proved to be closed, extending the GMP algebra initially derived for $\mC=1$ LLs to generic $\mC\geq1$ ideal flatbands. The closeness of the density algebra directly leads to the exact FCIs stabilized by generic $M-$body repulsive interactions which are Halperin type when  $\mC>1$ and typically non-Abelian when $M>2$. Thus the ideal condition Eq.~(\ref{defidealcond}) provides a general and exact statement for stabilizing FCIs regardless of the nonuniformness of Berry curvature. 

The paper is structured as follows. In Sec.~\ref{sec:motivation_summary}, we pose the problem by showing unusual numerical observations. In Sec.~\ref{sec:analytical}, we review key analytical results for ideal flatbands, emphasizing their real and momentum-space boundary conditions and define the emergent Hilbert space. We show in Sec.~\ref{sec:guidingcenter} a well defined guiding center in the ideal flatband problems which leads to the derivation of single-particle wavefunctions in Sec.~\ref{sec:idealwavefunction}. In Sec.~\ref{sec:densityalgebra}, the density algebra is derived and the origin of model FCIs is explained. We discuss application of the theory to FCIs and symmetry breaking phases~\cite{FCI_TBG_exp,Young_partiallyfilledTBG} in \mr flatband systems in Sec.~\ref{sec:conclusion}.

\section{Motivation from unusual numerical observations\label{sec:motivation_summary}}
The ideal flatbands are not abstract concepts, rather they can be realized in concrete models including Dirac fermion based models~\cite{Grisha_TBG,Jie_hierarchyidealband,Eslam_highC_idealband,Liang_DiracNonuniformB} and Kapit Mueller models~\cite{Kapit_Mueller,ModelFCI_Zhao,Dong_Mueller_20,Emil_constantBerry}. In particular, in the chiral twisted multilayer graphene models~\cite{Jie_hierarchyidealband,Eslam_highC_idealband} ideal flatbands of arbitrary $\mC\neq0$ are realized exactly. In this section, we use this model as a concrete platform to numerically study the interacting spectra, which point out a couple of important and unusual aspects of the projected density operator that motivates the theory to be discussed in the following sections.

\begin{figure}
\centering
\includegraphics[width=\linewidth]{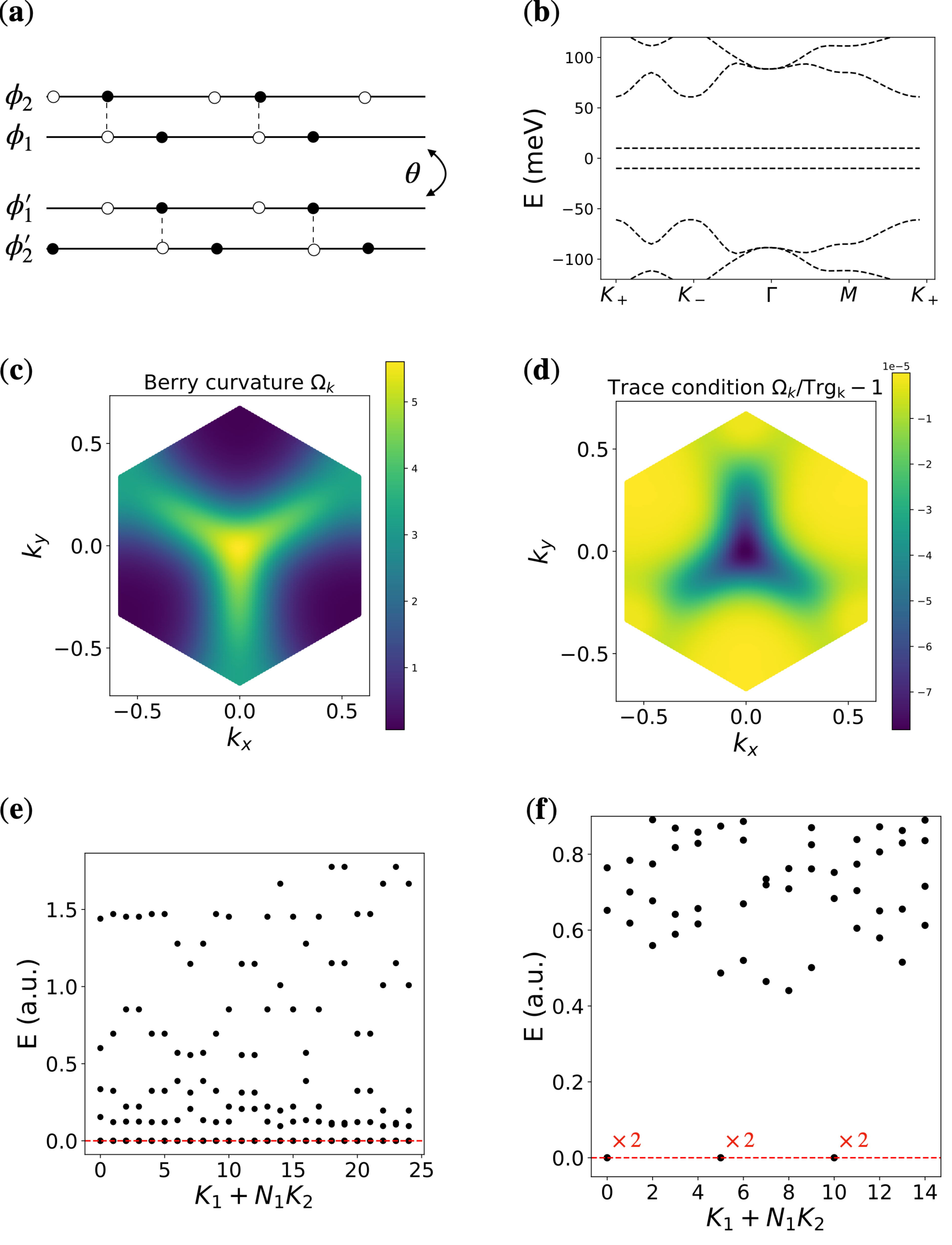}
\caption{(a) Illustration of the chiral twisted multilayered graphene model, where solid and empty dots represent the A and B sublattice, respectively, and the interlayer tunneling is assumed to be chiral and represented by the dashed line. The $\phi_{1,...,n}$ and $\phi'_{1,...,n}$ label the layer component. (b) The band structure where a tiny sublattice bias potential is used to split the two-fold degeneracy of the flatbands. The resulting two bands are entirely sublattice polarized. (c) and (d) The Berry curvature and the trace condition of $u^{\rm Bloch}_{\bm k}$ defined in the main text. The Berry curvature is positive definite. (e) The energy spectrum of the two-body $v_{m=1}$ interaction for two fermions on a $N_1 \times N_2=5\times5$ lattice. At each momentum $\bm k$, there are precisely $(m+1)\mC=4$ non-zero eigenvalues. (f) The energy spectrum of the three-body onsite interaction for $N=10$ bosons on a $N_1 \times N_2=5\times3$ lattice. The exact six-fold degenerate zero-energy ground states are non-Abelian model FCIs which are analogous to the FQH non-Abelian spin singlet state (NASS). In (a)-(f) we use $n=2$, namely the twisted double bilayer graphene model, as an example.}\label{fig_model}
\end{figure}

For self-consistency, we first briefly review the chiral twisted multilayer graphene models~\cite{Jie_hierarchyidealband,Eslam_highC_idealband}, although the model details are not crucial for the current discussion. Details about the chiral twisted graphene model and beyond can be found in the literature~\cite{Grisha_TBG,Grisha_TBG2,JieWang_NodalStructure,RafeiRen_TBG,Gerardo21PRB,Gerardo22,popov2020hidden,Oscar_hiddensym}. The chiral twisted multilayer model consists of two sheets of graphene multilayers each of which has $n-$layers and is A/B stacked without twist~\cite{PhysRevLett.106.156801,Geisenhof:2021aa,Senthil_NearlyFlatBand}. The top sheets are twisted relative to the bottom sheets as a whole by a twist angle $\theta$. See Fig.~\ref{fig_model}(a) for illustration of the setup. At magic twist angles, the model exhibits two exactly dispersionless degenerate bands showing in Fig.~\ref{fig_model}(b), which have Chern number $\mC=\pm n$. Such degeneracy can be easily lifted by sublattice contrasting potential, and we focus on the $\mC=n$ band. The flatband wavefunction is derived in Refs.~\cite{Jie_hierarchyidealband,Eslam_highC_idealband}, and its components on the outmost layers are found to be responsible for the high Chern number and zero-mode FCIs. We hence project the wavefunction to the outmost layers to get an effective single-component wavefunction, denoted as $\phi^{\rm Bloch}_{\bm k}(\bm r)$. It can be numerically verified that $\phi^{\rm Bloch}_{\bm k}(\bm r)$ is smooth in both $\bm k$ and $\bm r$. The Berry curvature associated to $\phi^{\rm Bloch}_{\bm k}(\bm r)$, defined through the standard definition,
\begin{equation}
    \Omega_{\bm k} = \epsilon_{ab}\partial_{\bm k}^a\bm A^b_{\bm k},\quad\bm A^a_{\bm k}\equiv-i\langle u_{\bm k}^{\rm Bloch}|\partial_{\bm k}^au^{\rm Bloch}_{\bm k}\rangle,
\end{equation}
is plotted in Fig.~\ref{fig_model}(c), which integrates to the Chern number $\mC=n$. Here $u^{\rm Bloch}_{\bm k}(\bm r) = e^{-i\bm k\cdot\bm r}\phi^{\rm Bloch}_{\bm k}(\bm r)$ is the cell-periodic part of the Bloch wavefunction. Its ideal geometry condition can be either analytically proved or numerically verified to be satisfied~\cite{Jie_hierarchyidealband,Eslam_highC_idealband}. See Fig.~\ref{fig_model} (d) for the plot of the trace condition for the $n=2$ model where errors are finite-size artifact that will vanish in the infinite-size limit.

We then consider interacting phenomena when the flatband is partially filled. We focus on finite systems on the torus geometry at band filling $\nu=N/(N_1 N_2)$, where $N$ is the number of particles and $N_{1,2}$ is the number of unit cells in each primitive direction of the lattice. The two-body density-density interaction takes the form of 
\begin{equation}
    H = \sum_{\bm q}v({\bm q}):\hat\rho_{\bm q}\hat\rho_{-\bm q}:,\label{defH}
\end{equation}
where $v({\bm q})$ is the Fourier transform of the interaction potential, $\hat\rho_{\bm q}$ is the band-projected density operator, and $: \ :$ is the normal ordering. Because of the translation invariance by lattice vectors, the energy spectrum of $H$ can be resolved by the center of mass momentum $(K_1,K_2)$. We are particularly interested in the repulsive interaction $v_m(\bm r_1-\bm r_2) = \sum_{\bm q}v_m(\bm q)e^{i\bm q\cdot(\bm r_1-\bm r_2)}$, whose Fourier transform $v_m(\bm q) $ has series expansion of orders no higher than $m$:
\begin{equation}
    v_m(\bm q) = c_0 + c_1|\bm q|^2 + c_2|\bm q|^{4} + ... + c_m|\bm q|^{2m}.\label{defVm}
\end{equation}
The above series expansion is equivalent as interaction range expansion, as the real space form of the $|\bm q|^{2n}$ term is given by the $2n_{\rm th}$ derivative of the contact interaction $\delta(\bm r_1-\bm r_2)$. In particular, the $v_0$ and $v_1$ are the shortest interactions for bosons and fermions, respectively \footnote{Interaction $v_m(\bm q) $ can be equivalently expanded in the Laguerre-Gaussian basis under which the coefficients are also called Haldane pseudopotentials~\cite{Haldane_hierarchy}. In the quantum Hall problems, Laguerre-Gaussian basis is preferred as it is the eigen-basis of two-particle coherent states~\cite{Haldane_hierarchy}. We comment that in flatband problems with nonuniform Berry curvature, the association of two-particle coherent states and Laguerre-Gaussian basis no longer exist, thereby there is no preference between the series expansion shown in Eq.~(\ref{defVm}) and the Haldane pseudopotential expansion as these two expansion bases are unitary related.}.

We first consider two particles interacting with $v_m$, where $m$ is even for bosons and odd for fermions. While it is unrealistic for bosons to occupy the flatband near charge neutrality of twisted double bilayer graphene, we can still use that setup to examine the band property. Remarkably, for any choice of $m$ and Chern number $\mathcal{C}$, the two-particle spectrum always shows a dispersive ``band'' with a fixed number of nonzero eigenvalues at each momentum $\bm k$, no matter what the lattice size is. The dimension of this finite-energy ``band'' is precisely $(m+1)\mC$ for fermions and $(m+1)\mC+1$ for bosons~\footnote{

This counting can be derived by first considering a $\mC-$layer FQH system and then mapping it to FCI. Suppose this $\mC-$layer FQH system is pierced by $N_{\phi}$ magnetic flux, such that there are $N_{\phi}$ sectors of the center-of-mass momentum. For bosons with the $v_{m=2n}$ interaction, they feel $n + 1$ Haldane pseudopotentials of even order, which affect both the intralayer and interlayer interactions, and $n$ Haldane pseudopotentials of odd order, which only affect the interlayer interaction. Because there are $\mC$ layers, each intralayer pseudopotential has $\mC$ copies and each interlayer pseudopotential has $\mC(\mC-1)/2$ copies. We know each copy contributes two nonzero energy levels in each of the $N_{\phi}$ FQH momentum sectors in the two-particle spectrum, so the total number of nonzero levels is $2(n+1)\left[\mC + \mC(\mC-1)/2\right] + 2n\mC(\mC-1)/2 = \mC\left[(2n+1)\mC+1\right]$. Mapping to FCIs, the number of momentum sectors are $\mC$ times larger, because we have $N_1N_2 = \mC N_{\phi}$. So the number of nonzero levels per FCI momentum sector is $\mC\left[(m+1)\mC+1\right] = (m+1)\mC+1$ for bosons. Similarly, we can get the counting for fermions with the $v_{m=2n+1}$ interaction as $(m+1)\mC$.}. The ``band'' dispersion reflects the fact that Berry curvature is nonuniform. Apart from these finite-energy levels, there are also massive zero modes whose dimension increases with the system size. The typical example with $m=1,\mathcal{C}=2$ is demonstrated in Fig.~\ref{fig_model}(e) for two fermions.  Such two-particle spectra are different from those in either LLs or generic flatbands. For generic flatbands without ideal geometry, the zero modes would not appear and the dimension of finite-energy levels is hence lattice-size dependent~\cite{hierarchy_FCI,zhao_review}. In a LL, as the $v_m$ interaction only picks out two-particle coherent states of relative angular momentum no greater than $m$, both a finite-energy band whose dimension is independent of the system size and massive zero modes appear, however, the ``band'' is non-dispersive due to the continuous translation symmetry. Therefore, Fig.~\ref{fig_model}(e) clearly shows there is an emergent projector property associated to the density operator of ideal flatbands which however, unlike in LLs, is influenced by the nonuniform Berry curvature.

We can further generalize Eq.~(\ref{defH}) to multibody interactions. In Fig.~\ref{fig_model}(f), the spectrum of the three-body onsite interaction is plotted for bosons at $\nu=2/3$ in $\mathcal{C}=2$ band. In this case, a six-fold exact ground-state degeneracy at zero energy is obtained. These zero modes are expected to be lattice analogs of the non-Abelian spin singlet states, which are generalizations of the Abelian Halperin topological order in bilayer LLL~\cite{Sterdyniak13}. This result indicates that suitable short-range $M$-body repulsions can stabilize model FCIs of both Abelian and non-Abelian types in high Chern number ideal flatbands. Moreover it points out that two-body interaction is not special: there should be a general property that the projected density operator obey in order to give rise to FCIs stabilized by generic $M$-body interactions~\footnote{Thorough this work we use the word ``exact FCIs'' and ``model FCIs'' interchangeably: the former refers to the fact that they are exact zero energy states for short-ranged interactions, and the latter is based on the observation that they also always have infinite large particle-cut entanglement spectra gap~\cite{PhysRevX.1.021014,Jie_hierarchyidealband}.}.

To summarize, numerical studies on interacting two-particle spectra and three-body interactions point out the importance of the flatband projected density operator which exhibits emergent projector properties and internal $\mC$ degrees of freedom playing the role of layers, although the underlying Berry curvature can be highly nonuniform. It motivates the following two questions:
\begin{itemize}
    \item What internal degrees of freedom in the scalar-valued wavefunction $u_{\bm k}(\bm r)$ play the role of layers?
    \item Why the nonuniform Berry curvature does not destabilize the Abelian and non-Abelian FCIs?
\end{itemize}

Thorough this work, we will show the answers to these two questions by deriving the general form of $\mC\neq0$ ideal flatband wavefunctions, illustrating their physical meanings and studying their density algebra which we find generalizes the GMP algebra Eq.~(\ref{projectedGMP}) in a nontrivial way.

\section{Analytical properties of ideal flatbands\label{sec:analytical}}
As we have shown, numerical experiments in ideal flatbands point out the unusual effects of ideal quantum geometry on interacting problems. To address the two questions at the end of the last section, we will focus on analytical derivations in the remaining parts of this article. We start with reviewing the ideal quantum geometric condition and emphasizing its relation to momentum-space K\"ahler condition in this section. The momentum-space K\"ahler condition gives novel properties of the Bloch wavefunction in many aspects, such as the unique boundary condition and the relation between Berry curvature and normalization factor. These general results were derived for ideal flatbands in Ref.~\cite{JieWang_exactlldescription}. In the next section Sec.~\ref{sec:guidingcenter}, we will discuss new results about their important implication for the well defined guiding center which establishes connection to LL physics in the presence of nonuniform Berry curvature.

\subsection{Ideal quantum geometry and K\"ahler geometry}
We first of all set up conventions in more detail. Our exact results applies to arbitrary system size $N_{1,2}$. The system is spanned by primitive lattice vectors $\bm a_{1,2}$ containing an area $|\bm a_1\times\bm a_2| = 2\pi S$ and throughout this work we set $S=1$ as the unit area scale. The reciprocal/ momentum space is spanned by reciprocal lattice vectors $\bm b_i$ such that $\bm a_i\cdot\bm b_j = 2\pi\delta_{ij}$ is satisfied.

We consider a single component complex valued Bloch wavefunction defined on this torus. Its cell periodic part $u^{\rm Bloch}_{\bm k}(\bm r)$ is lattice translational invariant following the Bloch theorem $u^{\rm Bloch}_{\bm k}(\bm r) = u^{\rm Bloch}_{\bm k}(\bm r+\bm a)$ for arbitrary lattice vectors $\bm a = m_1\bm a_1+m_2\bm a_2$ and $m_{1,2}\in\mathbb{Z}$. The $u^{\rm Bloch}_{\bm k}(\bm r)$ is assumed to be smooth in both $\bm k$ and $\bm r$. Momentum quantization restricts $\bm k = (n_1/N_1+\phi_1/2\pi)\bm b_1 + (n_2/N_2+\phi_2/2\pi)\bm b_2$ where $n_{1,2}\in\mathbb{Z}$ and $\phi_{1,2}$ are fictitious boundary condition fluxes. Our result applies to arbitrary $\phi_{1,2}$ too so without loss of generality we set $\phi_{1,2} = 0$. Because the flux space (torus formed by continuous variable $\phi_{1,2}$) is equivalent to the momentum space in the presence of translational symmetry, we use these two spaces interchangeably. The discussions in this work can be generalized to flux space for systems without translation symmetry where the notion of momentum is not defined.

The $u^{\rm Bloch}_{\bm k}(\bm r)$ satisfies the ideal quantum geometric condition while not necessarily carrying flat Berry curvature. We assume $u^{\rm Bloch}_{\bm k}$ has positive Chern number; negative Chern number is a simple generalization. As being a positive integer, the Chern number can be factorized into two positive integers $\mC=\mC_1\mC_2$ and the factorization, as we will discuss, is a gauge choice.

Systematic studies of flatbands with momentum-space holomorphicity start from Refs.~\cite{Martin_PositionMomentumDuality,Lee_engineering} where the authors found that the ideal condition is automatically satisfied if the cell-periodic part of Bloch wavefunction $u^{\rm Bloch}_{\bm k}(\bm r)$ is given by Eq.~(\ref{defukholo}). Very recently, Refs.~\cite{kahlerband1,kahlerband2,kahlerband3} pointed out the inverse is also true: the ideal condition generally implies  $u^{\rm Bloch}_{\bm k}$ can be written as Eq.~(\ref{defukholo}) under a gauge choice up to a positive-definite normalization factor $N_{\bm k}$:
\begin{equation}
    u^{\rm Bloch}_{\bm k}(\bm r) = N_{\bm k}u^{\rm holo}_k(\bm r),\quad\bar\partial_k u^{\rm holo}_k = 0,\label{defukholo}
\end{equation}
where the holomorphic coordinate $k$ (unbold letter) is defined as $k\equiv\omega^a\bm k_a$ and the associated anti-holomorphic derivative is defined as $\bar\partial_k \equiv \omega_a\partial_{\bm k}^a$. Technically, the notion of holomorphicity $\omega^a$ is defined from factorizing the constant uni-modular matrix $\omega^{ab}$ and the anti-symmetric tensor $\epsilon^{ab}$ as follows~\cite{Haldanegeometry}:
\begin{eqnarray}
    \omega^{ab} &=& \omega^{a}\omega^{b*} + \omega^{a*}\omega^{b},\nonumber\\
    i\epsilon^{ab} &=& \omega^{a*}\omega^{b} - \omega^{a}\omega^{b*}.\label{def_factorization}
\end{eqnarray}
The vectors satisfy $\omega^{a*}\omega_a = 1$, $\omega^a\omega_a = 0$ where and throughout the paper Einstein summation is implicitly assumed, and their indices are raised or lowered by $\omega^{ab}$. Note that in the isotropic case $(\omega^x, \omega^y) = (1,i)/\sqrt2$ the complex coordinate reduces to the conventional form $k = (k_x+ik_y)/\sqrt2$. We will work in the general case without assuming isotropy.

It is worth to mention that $\omega^{a}$ has an equivalent canonical definition as being the constant null vector of the quantum geometric tensor~\cite{JieWang_exactlldescription}:
\begin{equation}
    \mathcal Q^{ab}_{\bm k}\omega_b = 0\quad\text{for}\quad \forall\bm k,
\end{equation}
where the quantum geometric tensor is defined by using the covariant derivative of wavefunctions $D^a_{\bm k}=\partial_{\bm k}^a - iA_{\bm k}^a$, whose real symmetric and imaginary anti-symmetric parts are the Fubini-Study metric and the Berry curvature, respectively:
\begin{equation}
    \mathcal Q^{ab}_{\bm k} \equiv \langle D^a_{\bm k}u_{\bm k}|D^b_{\bm k}u_{\bm k}\rangle = g^{ab}_{\bm k} + i\frac{\epsilon^{ab}}{2}\Omega_{\bm k}.
\end{equation}

\subsection{Implications from K\"ahler geometry}
We have reviewed the relation between ideal quantum geometric condition and momentum-space holomorphicity~\cite{Martin_PositionMomentumDuality,kahlerband1,kahlerband2,kahlerband3}. In this section we discuss two implications from K\"ahler geometry: the uniqueness of the boundary condition and the relation between normalization factor and Berry curvature. They are important in deriving ideal flatband wavefunction and their density algebra.

\subsubsection{Uniqueness of the boundary condition}
Following Ref.~\cite{JieWang_exactlldescription}, we can generally denote the momentum-space boundary boundary condition of $u^{\rm holo}_k$ as:
\begin{eqnarray}
u^{\rm holo}_{k + b_i}(\bm r) &=& u^{\rm holo}_{ k}(\bm r)\cdot\exp\left(-i\bm b_i\cdot\bm r+i\phi_{k,  b_i}\right),\label{defkbc}
\end{eqnarray}
where the complex phase $\phi_{k,b}$ is not only required to be holomorphic in $k$ but also is constrained to satisfy a co-cycle relation~\cite{JieWang_exactlldescription},
\begin{equation}
-2\pi\mC = \phi_{b_1, b_2}-\phi_{0, b_2}+\phi_{0,  b_1}-\phi_{b_2, b_1}.\label{defCphi}
\end{equation}
Such a constraint Eq.~(\ref{defCphi}) is required from the simplified Chern number formula derived for ideal flatbands~\cite{JieWang_exactlldescription}:
\begin{equation}
    \mC = \frac{1}{2\pi i}\oint dk~\partial_k\ln u^{\rm holo}_{k}(\bm r),\label{Chernnumberform}
\end{equation}
which means that the winding of the wavefunction in the momentum space around the Brillouin zone at any fixed $\bm r$ must reflect the topology of the wavefunction. The co-cycle relation Eq.~(\ref{defCphi}) implies the boundary condition factor $\phi_{k,b}$ must be a linear function of holomorphic coordinate $k$~\cite{JieWang_exactlldescription}. The phase $\phi_{k,b}$ is the so-called factor of automorphy which appears in classifying holomorphic line bundles~\cite{Tatatheta1}. Without loss of generality, we choose the symmetric gauge,
\begin{equation}
    \phi_{k,b_j} = \mC b_j^*\left(-ik-ib_j/2\right) + \mC_j\pi,\quad j=1,2,\label{choicephikb}
\end{equation}
where $\mC_{1,2}$ are the two positive integers dividing the Chern number $\mC=\mC_1\mC_2$, and the ambiguity in their choice will prove to be unimportant.

\subsubsection{K\"ahler potential\label{sec:analytical:kahler}}
The anti-holomorphic component of the Berry connection $\bar A_{\bm k} \equiv \omega_a \bm A^a_{\bm k}$ can be computed from the standard definition:
\begin{eqnarray}
    \bar A_k &\equiv& -i\langle u^{\rm Bloch}_{\bm k}|\bar\partial_ku^{\rm Bloch}_{\bm k}\rangle,\nonumber\\
    &=& -i\int d^2\bm r~N_{\bm k}u^{\rm holo*}_k(\bm r)\bar\partial_k\left[N_{\bm k}u^{\rm holo}_k(\bm r)\right],\nonumber\\
    &=& -i\left(N^{-1}_{\bm k}\bar\partial_kN_{\bm k}\right)\int d^2\bm r~N^2_{\bm k}u^{\rm holo*}_k(\bm r)u^{\rm holo}_k(\bm r),\nonumber\\
    &=& -i\bar\partial_k\log N_{\bm k}.\label{def_barAk}
\end{eqnarray}
Then from $\Omega_{\bm k} = -i\left(\partial_k\bar A_{\bm k} - \bar\partial_kA_{\bm k}\right)$ where $A_{\bm k}$ is the complex-conjugate of $\bar A_{\bm k}$, we obtain the expression for the Berry curvature~\cite{JieWang_exactlldescription,Douglas:2009aa,semyon_largeN16}:
\begin{equation}
    \Omega_{\bm k} = -2\partial_k\bar\partial_k\ln N_{\bm k}.\label{BerryNorm}
\end{equation}

\begin{figure}
\centering
\includegraphics[width=\linewidth]{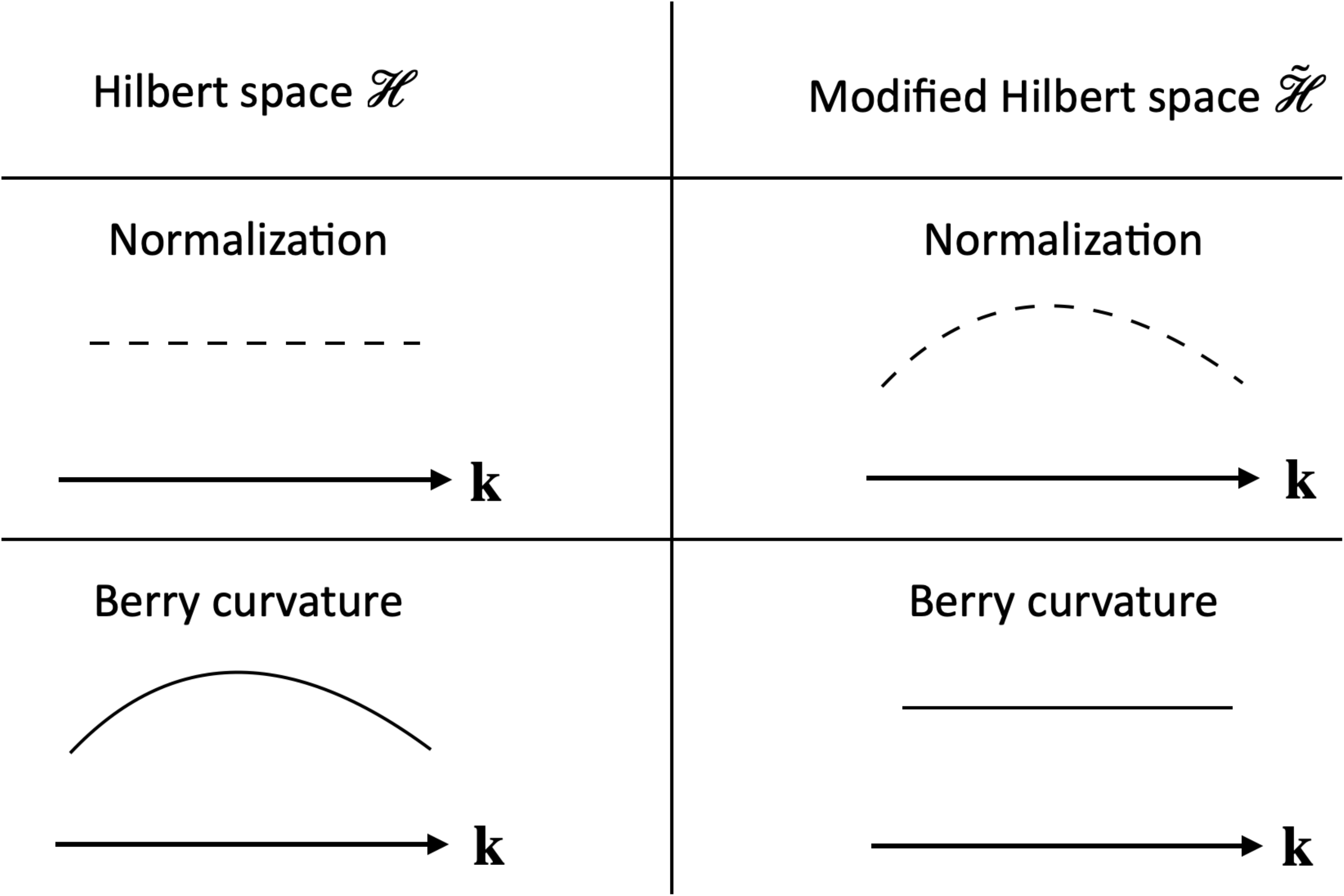}
\caption{The normalized flatband wavefunction defines a Hilbert space $\mathcal H$ where the Berry curvature is nonuniform. For ideal flatbands, the wavefunction's normalization can be tuned such that the resulting wavefunction $u_{\bm k}(\bm r)$ define a modified Hilbert space $\tmH$ in which Berry curvature can be exactly flattened. As a consequence $u_{\bm k}\in\tmH$ perceives identical $\bm k$-dependence as LLL wavefunctions. Such property greatly simplifies the problem, enabling a thorough characterization of flatband wavefunction from a momentum-space formulation. As an illustration, we used $\bm k$ to represent the 2D momentum-space. The normalization and Berry curvature are represented by the dashed and solid lines, respectively.}\label{fig_norm}
\end{figure}

Eq.~(\ref{BerryNorm}) has an important implication: for ideal flatbands, the Berry curvature can be effectively flattened by adjusted by the normalization of the wavefunction. More precisely, by factorizing $N_{\bm k}$ into,
\begin{equation}
    N_{\bm k} = e^{-\frac{\mC}{4}|\bm k|^2}\tilde N_{\bm k},
\end{equation}
the Berry curvature is accordingly split into $\Omega_{\bm k} = \mC + \tilde\Omega_{\bm k}$, {\it i.e.} the uniform and fluctuating part. The fluctuating part $\tilde\Omega_{\bm k} = -2\partial_k\bar\partial_k\ln\tilde N_{\bm k}$ averages to zero when being integrated over the Brillouin zone which does not contributes to the quantization.

We can thus flatten the Berry curvature by adjusting the normalization of the wavefunction. See Fig.~\ref{fig_norm} for an illustration. We define the following unnormalized wavefunction $u_{\bm k}(\bm r)$, which perceives constant Berry curvature, to be in the modified Hilbert space $\tmH$~\footnote{This is equivalent as changing the definition of the $L^2$ norm of the wavefunction so that it varies with $\bm k$: $\langle .|.\rangle'_{L^2} = \tilde N_{\bm k}^2\langle .|.\rangle_{L^2}$. The wavefunctions $u_{\bm k}(\bm r)$ normalized with the modified $L^2$ belong to the Hilbert space that we denote as $\tmH$.}:
\begin{equation}
    u_{\bm k}(\bm r) \equiv \tilde N_{\bm k}^{-1}u^{\rm Bloch}_{\bm k}(\bm r) = e^{-\frac{\mC}{4}\bm k^2}u^{\rm holo}_{k}(\bm r).\label{defuk}
\end{equation}

An important consequence is that $u_{\bm k}$ perceives a flattened Berry curvature and has identical $\bm k$-dependence as the LLL wavefunctions. Therefore the momentum-space formulation of the LLL problem can be directly applied to flatband problems which we will discuss in detail in the following section.

To summarize in this section we reviewed two implications from K\"ahler geometry, the uniqueness of the boundary condition and the relation between normalization and Berry curvature. We use the latter to define a modified Hilbert space $\tmH$ in which the wavefunctions $u_{\bm k}(\bm r)$ perceive constant Berry curvature. We summarize the real-space and boundary-condition space of $u_{\bm k}(\bm r)$ in below, by combining results from Eq.~(\ref{defkbc}), Eq.~(\ref{choicephikb}) and Eq.~(\ref{defuk}):
\begin{eqnarray}
    u_{\bm k}(\bm r+\bm a_i) &=& u_{\bm k}(\bm r),\label{def_rbc_uk}\\
    e^{-\frac{i\mC}{2}\bm b_i\times\bm k}u_{\bm k+\bm b_i}(\bm r) &=& (-1)^{\mC_i}e^{-i\bm b_i\cdot\bm r}\times u_{\bm k}(\bm r).\label{def_kbc_uk}
\end{eqnarray}
In Sec.~\ref{sec:idealwavefunction}, we will use these boundary conditions to determine the most general form of $u_{\bm k}(\bm r)$. In the next section Sec.~\ref{sec:guidingcenter}, we discuss the guiding centers in ideal flatbands and point out their simple algebra in $\tmH$, which is essential to many aspects of ideal flatbands.

\section{Guiding center in ideal flatbands\label{sec:guidingcenter}}
The guiding centers are particles' projected coordinates which typically are non-commutative. They and their algebra are known to play a crucial role in quantum Hall physics. In this section, we first review guiding center in LLs, followed by discussing the difficulty faced by generic flatbands. In the end, we show the ideal flatbands are exceptions, because there are well defined guiding centers with closed simple algebra.

\subsection{Guiding center in Landau levels}
In LLs, an electron's coordinate is split into the guiding center $\hat{\bm R}$ and Landau orbital $\hat{\bar{\bm R}}$ which correspond to the center and cyclotron motion of classical orbitals, respectively. Without loss of generality we choose a symmetric gauge under which the guiding center and Landau orbital are,
\begin{eqnarray}
    \hat R^a &=& -il_B^2\epsilon^{ab}\partial_b + r^a/2,\label{def_LLGC}\\
    \hat{\bar R}^a &=& +il_B^2\epsilon^{ab}\partial_b + r^a/2.
\end{eqnarray}
They form independent two sets of Heisenberg algebras $[\hat R^a, \hat{\bar R}^b] = 0$,
\begin{equation}
    ~[\hat R^a, \hat R^b] = -il_B^2\epsilon^{ab},\label{def_LLGCalgebra}
\end{equation}
and $[\hat{\bar R}^a, \hat{\bar R}^b] = +il_B^2\epsilon^{ab}$. Particularly Eq.~(\ref{def_LLGCalgebra}) can be called as ``guiding center algebra''.

The guiding centers and their algebra are one of the most important intrinsic features for physics insides a single LL~\cite{Haldanegeometry}. They are the essence of many ``ideal aspects'' of Landau levels: for instance they are the physical origin of the holomorphicity of LLs~\cite{Haldanegeometry}, they imply an infinity degrees of deformation symmetry~\cite{Winfinty_Cappelli,Winfinty_Karabali} and they are crucial to the stability of FQH states~\cite{MoreReadState,Haldane_hierarchy}.

A physically intuitive way to think of Eq.~(\ref{def_LLGCalgebra}) is based on the coherent state: the projected particle is no longer a point particle but has finite support and Eq.~(\ref{def_LLGCalgebra}) describes the uncertainty of its 2D coordinates. Such a coherent state can be translated or deformed while preserving the area, and in fact these processes are all generated by operations constructed from guiding centers. More precisely, the $\hat{\bm R}$ itself is the generator of magnetic translation group, and $\hat{\bm\Lambda}^{ab} = \{\hat R^a, \hat R^b\}/2$ deforms the metric of the coherent state while preserving its determinant, and in fact such metric deformation gives Hall viscosity response~\cite{Haldanegeometry,Read_viscosity09,Read_viscosity11,Grovmov_Son_PRX17}. Most generally, all area preserving deformation of Landau levels are generated by $e^{i\bm q\cdot\hat{\bm R}}$ whose leading and second order expansion in $\bm q$ correspond to the magnetic translation and metric deformation, respectively, and generally all higher order deformations exist. Due to the simple form of the guiding center algebra Eq.~(\ref{def_LLGCalgebra}), it is straightforward to derive the algebra for $e^{i\bm q\cdot\hat{\bm R}}$:
\begin{equation}
    ~[e^{i\bm q_1\cdot\hat{\bm R}}, e^{i\bm q_2\cdot\hat{\bm R}}] = 2i\sin\left(\frac{\bm q_1\times\bm q_2}{2}l_B^2\right)e^{i(\bm q_1+\bm q_2)\cdot\hat{\bm R}},\label{GMP_GC}
\end{equation}
which in fact is precisely equivalent to the GMP algebra shown in Eq.~(\ref{projectedGMP}) when noticing that the LLL projected density operator is $\hat\rho_{\bm q} = e^{i\bm q\cdot\hat{\bm R}}e^{-\frac{1}{4}\bm q^2l_B^2}$ where the Gaussian factor is the form factor.

\subsection{Why GMP algebra fails in generic flatbands}
Following E. Blount~\cite{BLOUNT1962305,Sundaram_Niu,Martin_PositionMomentumDuality,Di_Review}, the projected coordinates $\hat{\bm r}$ are given in below for band systems:
\begin{eqnarray}
    \hat Q^{a}_{\rm Bloch} &\equiv& -i\partial_{\bm k}^a - \langle u^{\rm Bloch}_{\bm k}|i\partial_{\bm k}^au^{\rm Bloch}_{\bm k}\rangle,\nonumber\\
    &=& -i\partial_{\bm k}^a + A_{\bm k}^a.\label{def_GCband}
\end{eqnarray}
To avoid the real-space representation as used for LLs, we denote the above $\bm k-$space representation as $\hat{\bm Q}$. It is straightforward to check their commutator is the Berry curvature two form,
\begin{eqnarray}
    ~[\hat Q^a_{\rm Bloch}, \hat Q^b_{\rm Bloch}] &=& -i\partial_{\bm k}^aA_{\bm k}^b - \left(a \leftrightarrow b\right),\nonumber\\
    &=& -i\epsilon^{ab}\Omega_{\bm k}.
\end{eqnarray}
This immediately means that higher-order commutators such as $[\hat Q^a_{\rm Bloch}, [\hat Q^b_{\rm Bloch}, \hat Q^c_{\rm Bloch}]]$ depend on the Berry curvature derivative $\partial_{\bm k}^a\Omega_{\bm k}$, which generally is non-vanishing in flatband systems. As a consequence, the GMP algebra is \emph{not} expected to exist, reducing the stability of correlated phases of matter due to such geometric instability~\cite{Jackson:2015aa,PhysRevB.90.075104,RoyPRB16}.

\subsection{Emergent guiding center algebra in ideal flatbands}
In this section, we point out the exact and extremely simple guiding center algebra emerges in ideal flatbands. The key reason lies in the K\"ahler potential discussed in Sec.~\ref{sec:analytical:kahler}. It means the complication from nonuniform part of Berry curvature can be exactly and completely removed by adjusting the normalization of the Bloch wavefunction.

The $\tmH-$space projected coordinate operator is,
\begin{eqnarray}
    \hat Q^a &=& -i\partial_{\bm k}^a - \langle u_{\bm k}|i\partial_{\bm k}^au_{\bm k}\rangle,\nonumber\\
    &=& -i\partial_{\bm k}^a - \mC \epsilon^{ab}k_b/2.\label{def_GCidealband}
\end{eqnarray}
Eq.~(\ref{def_GCidealband}) is derived as follows: we can first look at $\omega_a\hat Q^a$; following the definition of projected coordinates Eq.~(\ref{def_GCband}) and wavefunction Eq.~(\ref{defuk}) we have:
\begin{eqnarray}
    \omega_a\hat Q^a &=& -i\bar\partial_k - \langle u_{\bm k}|i\bar\partial_k u_{\bm k}\rangle,\nonumber\\
    &=& -i\bar\partial_k + i\mC k/2,
\end{eqnarray}
which immediately implies Eq.~(\ref{def_GCidealband}), because $\hat{ Q}^a$ is the linear combination of $\omega_a\hat Q^a$ and its complex conjugate. Technically we have treated $u_{\bm k}(\bm r)$ as normalized wavefunctions when computing expectation value of any operator $\hat O$ in this way $O_{\bm k\bm k'} \equiv \int d^2\bm r~u^*_{\bm k}(\bm r)O(\bm r)u_{\bm k'}(\bm r)$ without dividing their normalization $\tilde N_{\bm k}\tilde N_{\bm k'}$.

Interestingly Eq.~(\ref{def_GCidealband}) takes a similar form as the guiding center of LLs Eq.~(\ref{def_LLGC}) but formulated in momentum-space and using an effective magnetic length $l_B^2=\mC$. In analogy to the Landau orbital operator in LLs, we define the following operator
\begin{equation}
    \hat{\bar Q}^a \equiv -i\partial_{\bm k}^a + \mC\epsilon^{ab}k_b/2,
\end{equation}
that commutes with $\hat Q^a$. It is easy to check that $\hat{\bm Q}$ and $\hat{\bar{\bm Q}}$ individually obey the Heisenberg algebra:
\begin{eqnarray}
    ~[\hat Q^a, \hat Q^b] &=& -i\epsilon^{ab}\mC,\\
    ~[\hat{\bar Q}^a, \hat{\bar Q}^b] &=& +i\epsilon^{ab}\mC. 
\end{eqnarray}

The $\hat{\bm Q}$ and $\hat{\bar{\bm Q}}$ are in fact the momentum-space form of guiding center and Landau orbital, respectively. For this reason we can call them as dual guiding center and dual Landau orbital. The existence of such well defined two sets of Heisenberg algebra is important to the physics in the ideal flatbands. For instance, the so-called ``LLL condition'', the magnetic translation algebra, pseudopotentials are all well defined in an exact manner for ideal flatbands, but generalized from LLs in a nontrivial way. We illustrate their meaning one by one thorough this paper. To begin with, we discuss the ``dual LLL condition'', which is seen from noticing that the modified Hilbert space $\tmH$ is completely annihilated by the following ladder operator analogous to the Landau level annihilation operator used in quantum Hall physics:
\begin{equation}
    \hat{\bar a} u_{\bm k} = 0,\quad \forall u_{\bm k}\in\tmH,
\end{equation}
where
\begin{equation}
    \hat{\bar a} \equiv \omega_a\hat{\bar Q}^a = -i\bar\partial_k + i\mC \bar k/2.
\end{equation}

Moreover, the dual guiding center $\bm Q$ leaves the Hilbert space $\tmH$ invariant, because it commutes with $\bar{\bm Q}$. Then there exists another set of ladder operators,
\begin{equation}
    \hat a \equiv \omega^*_a\hat Q^a/\sqrt{\mC},\quad \hat a^\dag \equiv \omega_a\hat Q^a/\sqrt{\mC},\quad [\hat a, \hat a^\dag] = 1, \label{defaadag}
\end{equation}
such that they map states within $\tmH$:
\begin{equation}
    \hat a u_{\bm k} \in \tmH,\quad \hat a^\dagger u_{\bm k} \in \tmH,\quad \forall u_{\bm k} \in \tmH.
\end{equation}
In this sense, $\tmH$ behaves like the LLL Hilbert space in the quantum Hall physics, and hence we expect $u_{\bm k}$ shows similarity in its form with the LLL wavefunction. In the next section, we utilize such properties to fully characterize the ideal flatband wavefunction, following the same idea how torus quantum Hall wavefunction was initially derived~\cite{haldanetorus1,haldanetorus2}.

We comment that the choice of gauges, {\it i.e.} the Brillouin zone factorization $\mC_{1,2}$ and the particular form of $\phi_{k,b}$, only affects the concrete form of dual guiding centers and dual Landau orbitals. The fact that for ideal flatbands the existence of such complete separation of two sets of Heisenberg algebra is a fundamental consequence of ideal quantum geometry and gauge independent. This is also the key for the existence of exact properties even in the presence of nontrivial Berry curvature.

To summarize this section, the flattening of Berry curvature in the modified Hilbert space $\tmH$ greatly simplifies the ideal flatband problem. The $\tmH-$space projected coordinates, {\it i.e.} the dual guiding centers, map states within $\tmH$ and obey an identical algebra as the guiding centers in LLs. Therefore, tools from quantum Hall physics can be applied to analyze ideal flatbands in many aspects which we discuss in following sections.


\section{Ideal flatband wavefunction\label{sec:idealwavefunction}}
The emergence of the guiding center and their exact algebra in ideal flatbands can greatly simplify the flatband problem while retaining their nontrivial aspect: nonuniform quantum geometries. It also motivates that tools from the quantum Hall physics can be applied to analyze the flatband problem. In this section, we point out it is crucial to adopt a position-momentum exchanged view to carefully compare the two system: ideal flatband and Landau levels~\cite{Martin_PositionMomentumDuality}. Importantly, we point out that the wavefunctions of ideal flatbands are highly constrained by their quantum geometry, exactly for the same reason how LLL wavefunction is highly constrained by the magnetic field~\cite{haldanetorus1,haldanetorus2}. Utilizing the emergent dual guiding center and their algebra, we define dual version of the magnetic translation group and use it to fully characterize the ideal flatband wavefunction. We prove there is a unique general form of the single-particle Bloch wavefunction for all topological ideal flatbands, and provide their explicit first quantized expression.

\subsection{Dual magnetic translation group}
In Landau levels, the guiding center $\hat{\bm R}$ generates the magnetic translation group (MTG), which adiabatically transports particles while preserving them within the LL they initially start with. In this section, we define the dual-MTG for ideal flatbands, such that this group adiabatically transports Bloch wavefunctions while preserving their ideal quantum geometric condition. We then use the dual-MTG to derive ideal flatband wavefunctions.

The dual-MTG is generated by the dual guiding centers. Its group elements are:
\begin{equation}
    t(\bm q) = e^{i\bm q\cdot\hat{\bm Q}} = e^{-\frac{i\mC}{2}\bm q\times\bm k}e^{\bm q\cdot\partial_{\bm k}}.\label{def_tq}
\end{equation}

As discussed in the last section, the dual guiding center leaves $\tmH$ invariant. Therefore $t(\bm q)$ also maps states within $\tmH$ and preserves their ideal quantum geometric condition. The algebra of the dual-MTG is easily derived based on the non-commutativity of $\bm Q$:
\begin{equation}
    t(\bm q_1)t(\bm q_2) = e^{i\mC\bm q_1\times\bm q_2}t(\bm q_2)t(\bm q_1).
\end{equation}

From the above we see translating particles across the whole Brillouin zone commute because $[t(\bm b_1), t(\bm b_2)] = 0$. However, they are not the minimal translations that commute with each other. The minimal commuting set is determined by fractional translations by distance $\tbm b_{1,2}$,
\begin{equation}
    \tbm b_1 = \bm b_1/\mC_1,\quad \tbm b_2 = \bm b_2/\mC_2.
\end{equation}
$\tbm b_{1,2}$ thus define a smaller Brillouin zone which is only $1/\mC$ fraction of the flatband Brillouin zone. Consequently, their real-space unit cell is enlarged by a factor of $\mC$. We denote the real space-lattice lattice vectors as,
\begin{equation}
    \tbm a_1 = \mC_1\bm a_1,\quad \tbm a_2 = \mC_2\bm a_2.
\end{equation}

\begin{figure}
\centering
\includegraphics[width=\linewidth]{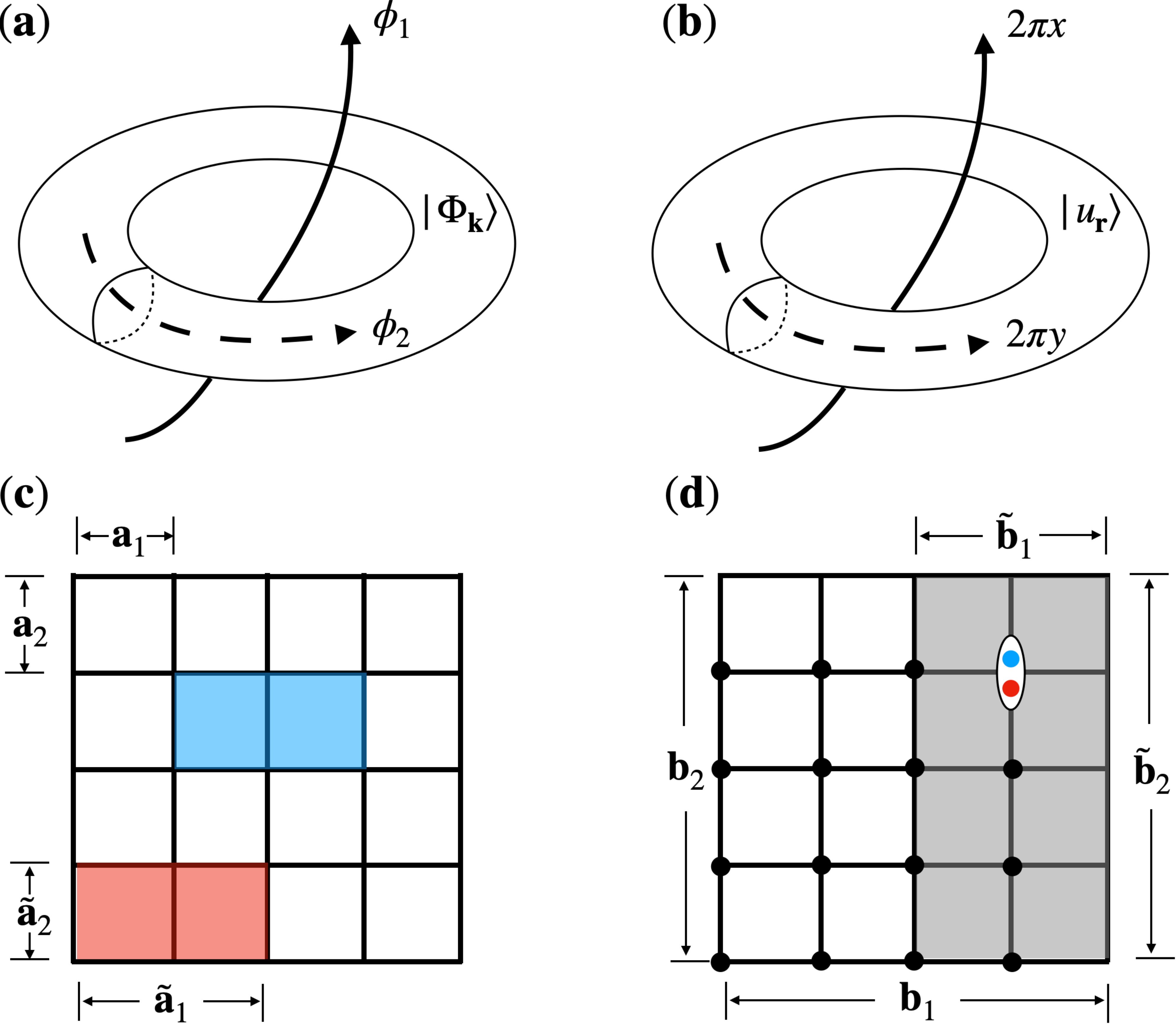}
\caption{(a) Inserting boundary condition fluxes $\phi_{1,2}$ pumps the momentum. Such flux insertion process is generated by the magnetic translation operator in quantum Hall systems. (b) In ideal flatbands, one can define a dual magnetic translation and use the coordinate $\bm r$ to pump states $|u_{\bm r}\rangle$. Here $\bm r = x\bm a_1 + y\bm a_2$ with $x,y\in[0,1)$ is defined within the unit cell spanned by $\bm a_{1,2}$. (c) Example of the lattice unit cells $\bm a_{1,2}$ and the enlarged unit cells $\tbm a_{1,2}$ on a $N_1=N_2=4$ system for Chern number $\mC=2$ ($\mC_1=2$, $\mC_2=1$). The blue and red squares are representative magnetic unit cells of two lowest Landau level state. (d) The momentum-space of the same system, where there are in total $N_{1}N_{2}=16$ degrees of freedom representing orthogonal $16$ flatband states denoted by the black dots. In mapping to the lowest Landau levels, the first Brillouin zone gets down-folded by $\mC$ times into the grey area spanned by $\tbm b_{1,2}$ but each momentum point is enriched by $\mC$ colors. The total degrees of freedom is unchanged.}\label{fig_BZ}
\end{figure}

We illustrate an example of the newly introduced lattice vectors with $\mC=2$, $\mC_1=2$, $\mC_2=1$ in Fig.~\ref{fig_BZ} (c) and (d). As we will see shortly, the new lattice vectors in fact define new degrees of freedom that are crucial in constructing the ideal flatband wavefunctions and in mapping flatband to LL physics. Before going into the details of the wavefunction, we first revisit the boundary condition Eq.~(\ref{def_kbc_uk}) which highlights the importance of position-momentum duality and sheds light into the connection between ideal flatband and LLs.

\subsection{Boundary condition revisited}
It is important to notice that the momentum-space boundary condition Eq.~(\ref{def_kbc_uk}) in fact can be precisely rewritten in terms of the dual MTG as follows:
\begin{equation}
    e^{i\bm b_i\cdot\hat{\bm Q}}u_{\bm k}(\bm r) = (-1)^{\mC_i}e^{-i\bm b_i\cdot\bm r}u_{\bm k}(\bm r).\label{dMTG2}
\end{equation}

Eq.~(\ref{dMTG2}) is akin to the boundary condition in quantum Hall problems. Recall that for quantum Hall problems defined on torus of length $\bm L_{i=1,2}$ consisting $N_{1,2}$ magnetic unit cells along the two directions, periodic magnetic translation across the torus imposes the following boundary condition for all quantum Hall states $\Phi_{\bm k}(\bm r)$:
\begin{equation}
    e^{i\bm L_i\times\hat{\bm R}/l_B^2}\Phi_{\bm k}(\bm r) = (-1)^{N_i}e^{i\phi_i}\Phi_{\bm k}(\bm r),\label{defmaxbc}
\end{equation}
where $\phi_{1,2}$ are the boundary condition fictitious fluxes, see Fig.~\ref{fig_BZ} (a) for example~\footnote{We write $e^{(\bm s \times \bm t)\cdot \hat{\bm z}}$ as $e^{\bm s \times \bm t}$ throughout the article for simplicity.}.

It is useful to compare Eq.~(\ref{dMTG2}) with Eq.~(\ref{defmaxbc}). First of all, it is interesting to notice that for ideal flatbands it is the wavefunction's coordinate $e^{-i\bm b_i\cdot\bm r}$ that tunes the boundary condition of the dual-MTG. Thereby tuning $\bm r$ generates a Thouless-pump~\cite{Thouless_1983} that transports states, in analogous to tuning $\phi_{1,2}$ in quantum Hall problems that transports particles. See Fig.~\ref{fig_BZ} (a) and (b) for comparison of the ``flux insertion process'' in Landau levels and ideal flatbands.

In fact, the dual Thouless Thouless pump can be numerically verified, for instance, by using the chiral twisted multilayer graphene model~\cite{Jie_hierarchyidealband,Eslam_highC_idealband}. In Fig.~\ref{fig_pattern_zero}, we plotted the absolute value of the ideal flatband wavefunction $u^{\rm Bloch}_{\bm k}(\bm r)$ from the $\mC=2, 3$ model for the top and bottom panel, respectively. We see at any fixed $\bm r$ the wavefunction exhibits $\mC$ zeros. Moreover, when we continuously vary $\bm r$, the pattern of zeros evolve analogous to charge pumping. There always exists a lattice vector $\bm a$ such that the indices of zeros get exchanged when $\bm r \rightarrow \bm r+\bm a$. Such zero exchange phenomenon will immediately become clear when we derive the wavefunction in the next section.

Lastly we comment that MTG and dual-MTG differ in many aspects, yet correspondences exist. A comparison between MTG and dual-MTG can be found in Table~\ref{MTGcomparison}. In particular, the Brillouin zone factorization $\mC_{1,2}$ in ideal flatbands corresponds to the torus factorization $N_{1,2}$ in Landau levels. This can be understood as follows. In quantum Hall setting, since the LLL wavefunctions are holomorphic in real-space coordinates and the space has topology of the torus, the wavefunctions have $N_{\phi} = N_1N_2$ zeros by the Riemann-Roch theorem. The latter relates the number of zeros to the dimension of the vector space of holomorphic sections of degree $N_{\phi}$ line bundle. In other words, there are in total $N_{\phi}$ independent wavefunctions on the LLL. Similar reasoning applies to ideal flatbands. Due to the Chern number formula Eq.~(\ref{Chernnumberform}), at any fixed $\bm r$ the ideal flatband wavefunction is a holomorphic section of degree $\mC$ line bundle, so it will have exactly $\mC$ zeros in the first Brillouin zone spanned by $\bm b_{1,2}$. This implies $\mC$ independent solutions satisfying the boundary condition Eq.~(\ref{def_kbc_uk}). In the next section we will construct an explicit basis of the wavefunctions.

\renewcommand{\arraystretch}{1.5} 
\begin{table*}[t]
\begin{tabular}{ |p{4.5cm}|p{6.5cm}|p{6.5cm}|}
    \hline
    & dual-MTG (for ideal flatbands): $e^{i\bm q\cdot\bm Q}$ & MTG (for LLs): $e^{i\bm d\times\bm R/l_B^2}$ \\
    \hline
    Torus &Brillouin zone; $\bm b_i = \mC_i\tbm b_i$; $\mC = \mC_1\times\mC_2$ & Real-space torus; $ \bm L_i = N_i\tbm a_i$;  $N_{\phi} = N_1\times N_2$\\
    \hline
    Unit cell & $\tbm b_1\times\tbm b_2 = 2\pi/(\mC S)$;\quad $\tbm a_1\times\tbm a_2 = 2\pi (\mC S)$ & $\tbm a_1\times\tbm a_2 = 2\pi l_B^2$;\quad $\tbm b_1\times\tbm b_2 = 2\pi/l_B^2$\\
    \hline
    Maximal boundary condition & $e^{i\bm b_i\cdot\hat{\bm Q}}|u_{\bm r}\rangle = (-1)^{\mC_i}e^{i\tilde\phi_i}|u_{\bm r}\rangle$ 
    &$e^{i\bm L_i\times\hat{\bm R}/l_B^2}|\Phi_{\bm k}\rangle = (-1)^{N_i}e^{i\phi_i}|\Phi_{\bm k}\rangle$\\
    \hline
    Minimal boundary condition & $e^{i\tbm b_i\cdot\hat{\bm Q}}|u_{\bm r}\rangle = -e^{-i\tbm b_i\cdot\bm r}|u_{\bm r}\rangle$ & $e^{i\tbm a_i\times\bm R/l_B^2}|\Phi_{\bm k}\rangle = -e^{i\tbm a_i\cdot\bm k}|\Phi_{\bm k}\rangle$\\
    \hline
    Representation & $|u_{\bm r}\rangle$ with $\bm r = \sum_{i=1,2}(n_i-\tilde\phi_i/2\pi)\tbm a_i/\mC_i$ & $|\Phi_{\bm k}\rangle$ with $\bm k = \sum_{i=1,2}(n_i+\phi_i/2\pi)\tbm b_i/N_i$\\
    \hline
    Transformation & $e^{i\bm q\cdot\hat{\bm Q}}|u_{\bm r}\rangle = e^{\frac{i}{2}\bm q\cdot\bm r}|u_{\bm r+\bm r_{\bm q}}\rangle$ & $e^{i\bm d\times\hat{\bm R}/l_B^2}|\Phi_{\bm k}\rangle = e^{\frac{i}{2}\bm d\cdot\bm k}|\Phi_{\bm k+\bm q_{\bm d}}\rangle $\\ 
    \hline
\end{tabular}
\caption{Comparison of the dual-MTG with the standard one requires a view from position-momentum duality. The flatband Brillouin zone corresponds to the entire real-space torus defining the quantum Hall problem, and fractional flatband Brillouin zone $\tbm b_{1,2}$ corresponds to one magnetic unit cell. There are $\mC$ dual magnetic unit cells in flatband problem, and $N_{\phi}$ magnetic unit cells in the quantum Hall problem. Interestingly, for ideal flatbands it is the coordinate $\bm r$, instead of momentum $\bm k$, that tunes the boundary condition and labels representations. The coordinate $\bm r$ induced charge pump is shown in Fig.~\ref{fig_pattern_zero}.}\label{MTGcomparison}
\end{table*}

\subsection{Irreducible representation of the dual magnetic translation group}
In the end of the last section, we have mentioned there must be $\mC$ linearly independent wavefunctions given the momentum-space boundary condition Eq.~(\ref{def_kbc_uk}) as a consequence of the Riemann-Roch theorem. In this section, we use the dual-MTG to derive them. Since wavefunctions are irreducible representations of the dual-MTG, we start with discussing the representation in general. It is useful to first review the representation in Landau level problems as the dual-MTG shares many similarities with the MTG although also quite different.

\subsubsection{Landau level and the usual magnetic translation group}
We denote the magnetic unit cell by $\tbm a_{1,2}$ too, and consider a system of length $\bm L_i = N_i\tbm a_i$ for $i=1,2$. It will be clear later why we use the same notation as flatbands. Quantum Hall wavefunctions are labeled by a magnetic momentum $\bm k$. Besides being eigenstate of MTG across the sample Eq.~(\ref{defmaxbc}), they are also eigenstates by translating $\tbm a_{i=1,2}$:
\begin{equation}
    e^{i\tbm a_i\times\bm R/l_B^2}|\Phi_{\bm k}\rangle = -e^{i\tbm a_i\cdot\bm k}|\Phi_{\bm k}\rangle.\label{defminbc}
\end{equation}

We can call Eq.~(\ref{defmaxbc}) and Eq.~(\ref{defminbc}) as the maximal boundary condition and the minimal boundary condition for the quantum Hall problem, respectively. These two boundary conditions in together quantize the momentum points onto lattice shifted by the boundary condition fictitious flux $\phi_{1,2}$:
\begin{equation}
    \bm k = \left(\frac{n_1}{N_1} + \frac{\phi_1}{2\pi}\right)\tbm b_1 + \left(\frac{n_2}{N_2} + \frac{\phi_2}{2\pi}\right)\tbm b_2,\label{quan_k}
\end{equation}
where $\tbm b_{1,2}$ are the lattice vectors reciprocal to $\tbm a_{1,2}$.

For a general vector $\bm d$ allowed by the boundary condition, state $|\Phi_{\bm k}\rangle$ transforms as,
\begin{eqnarray}
    e^{i\bm d\times\bm R/l_B^2}|\Phi_{\bm k}\rangle &=& e^{\frac{i}{2}\bm d\cdot\bm k}|\Phi_{\bm k+\bm q_{\bm d}}\rangle,\label{generalMTG}\\
    \left(\bm q_{\bm d}\right)_a &\equiv& l_B^2\epsilon_{ab}\bm d^b.
\end{eqnarray}
It is straightforward to verify that Eq.~(\ref{generalMTG}) preserves the magnetic translation algebra.

The maximal and minimal boundary conditions Eq.~(\ref{defmaxbc}) and Eq.~(\ref{defminbc}) uniquely determines the expression of the LLL wavefunction which is holomorphic in real-space coordinates up to a Gaussian factor~\cite{haldanetorus1,haldanetorus2}. The first quantized wavefunction can be written by the Weierstrass sigma function $\sigma(z)$,
\begin{eqnarray}
    \Phi_{\bm k}(\bm r) &\equiv& \langle\bm r|\Phi_{\bm k}\rangle,\label{defsigmawf}\\
    &=& \sigma(z-z_k)e^{z^*_k z/l_B^2}e^{-\frac12|z_k|^2/l_B^2}e^{-\frac12|z|^2/l_B^2},\nonumber
\end{eqnarray}
where $z_k = -ikl_B^2$ and $k=\omega^a\bm k_a$ is the complex momentum coordinate. The Weierstrass sigma function satisfies a quasi-periodic translation property~\cite{haldanemodularinv,haldaneholomorphic,Jie_MonteCarlo,JieWang_NodalStructure}:
\begin{equation}
    \sigma(z+\tilde a_{i}) = -e^{\tilde a^{*}_i(z + \tilde a_i/2)/l_B^2}\sigma(z),\quad i=1,2.\label{defsigmaperiod}
\end{equation}

\subsubsection{Basis functions of ideal flatbands and the dual magnetic translation group\label{sec:sub:basis}}
Having reviewed the MTG properties for Landau level states, we now proceed to discuss ideal flatband wavefunctions. We will derive the representation for the dual-MTG in complete analogy to the derivation from Eq.~(\ref{defminbc}) to Eq.~(\ref{defsigmawf}) used for Landau levels.

The fact that real-space coordinate $\bm r$ tunes the boundary condition of the dual-MTG motivates us to interpret the ideal flatband wavefunction as the projection of a quantum state $|u_{\bm r}\rangle$ into the momentum coordinate space, in complete analogy to quantum Hall wavefunctions being understood as state $\Phi_{\bm k}$ labeled by boundary condition $\bm k$ projected into the real-space as shown in Eq.~(\ref{defsigmawf}):
\begin{equation}
    u_{\bm k}(\bm r) = \langle\bm k|u_{\bm r}\rangle.
\end{equation}

Similar to LL wavefunctions, the ideal flatband states are not only constrained by the maximal boundary condition Eq.~(\ref{dMTG2}) but by a minimal boundary condition defined in below:
\begin{equation}
    e^{i\tbm b_i\cdot\bm Q}|u_{\bm r}\rangle = -e^{-i\tbm b_i\cdot\bm r}|u_{\bm r}\rangle.\label{dMTG3}
\end{equation}

The two boundary conditions Eq.~(\ref{dMTG2}) and Eq.~(\ref{dMTG3}) quantize the coordinates $\bm r$ on a lattice,
\begin{equation}
    \bm r = \left(\frac{m}{\mC_1} - \frac{\tilde\phi_1}{2\pi}\right)\tbm a_1 + \left(\frac{n}{\mC_2} - \frac{\tilde\phi_2}{2\pi}\right)\tbm a_2,\label{quan_r}
\end{equation}
where $\tilde\phi_i$ satisfying $e^{-i\tilde\phi_i} = e^{i\bm b_i\cdot\bm r}$ is precisely the boundary condition for the dual-MTG. For a generic vector $\bm q$, the state $|u_{\bm r}\rangle$ transforms as:
\begin{eqnarray}
    e^{i\bm q\cdot\bm Q}|u_{\bm r}\rangle &=& e^{\frac{i}{2}\bm q\cdot\bm r}|u_{\bm r+\bm r_{\bm q}}\rangle,\label{dMTG1}\\
    \left(\bm r_{\bm q}\right)^a &=& \mC\epsilon^{ab}\bm q_b.
\end{eqnarray}
We remind the reader that $\mC$ above should be understood as $\mC S$ since we have set the unit cell area $2\pi S = 2\pi$ throughout the paper. In this way, $\mC S$ has the dimension of area to convert momentum into coordinate.

The quantization of $\bm r$ in Eq.~(\ref{quan_r}) implies $\mC$ independent states which are related by lattice translations. Denoting,
\begin{equation}
    \bm a_{\bm\sigma} \equiv \sigma_1\bm a_1 + \sigma_2\bm a_2
\end{equation}
the first quantized wavefunctions of the $\mC$ independent states can be labeled as,
\begin{equation}
    v^{\bm\sigma}_{\bm k}(\bm r) \equiv \langle\bm k|u_{\bm r+\bm a_{\bm\sigma}}\rangle,\label{defvsigma}\\
\end{equation}
where $\bm\sigma=\left(\sigma_1,\sigma_2\right)$ is the integer-valued color-index with $\sigma_i\in[0,\mC_i-1]$.

For notational simplicity, we particularly drop the color index for the $\bm\sigma=(0,0)$ component. Its wavefunction is:
\begin{equation}
    v_{\bm k}(\bm r) = \sigma(z-z_k)e^{\frac{1}{\mC}\bar zz_k}e^{-\frac{1}{2\mC}|z|^2}e^{-\frac{1}{2\mC}|z_k|^2},\label{defv}
\end{equation}
where $z_k = -i\mC k$. According to Eq.~(\ref{defvsigma}), wavefunctions of other colors are obtained from lattice translations:
\begin{equation}
    v^{\bm\sigma}_{\bm k}(\bm r) = v_{\bm k}(\bm r + \bm a_{\bm\sigma}).\label{defvsigma1stquantized}
\end{equation}

It is worth to notice that $e^{i\bm k\cdot\bm r}v_{\bm k}(\bm r)$ is precisely the LLL wavefunction Eq.~(\ref{defsigmawf}) of magnetic length $l_B=\sqrt{\mC}$. Wavefunctions with other colors are distinguished by the their minimal boundary condition which we discuss later in the section of mapping flatband to LLs. We conclude this section by commenting that: all $v^{\bm\sigma}_{\bm k}(\bm r)$ obey the required momentum-space boundary condition Eq.~(\ref{def_kbc_uk}), which can be straightforwardly verified by using the quasi-periodicity of the sigma function. Thereby $v^{\bm\sigma}_{\bm k}(\bm r)$ are the basis functions spanning the $\mC-$dimensional holomorphic line bundle on the momentum-space manifold for any fixed coordinate $\bm r$.

\begin{figure*}
\centering
\includegraphics[width=\linewidth]{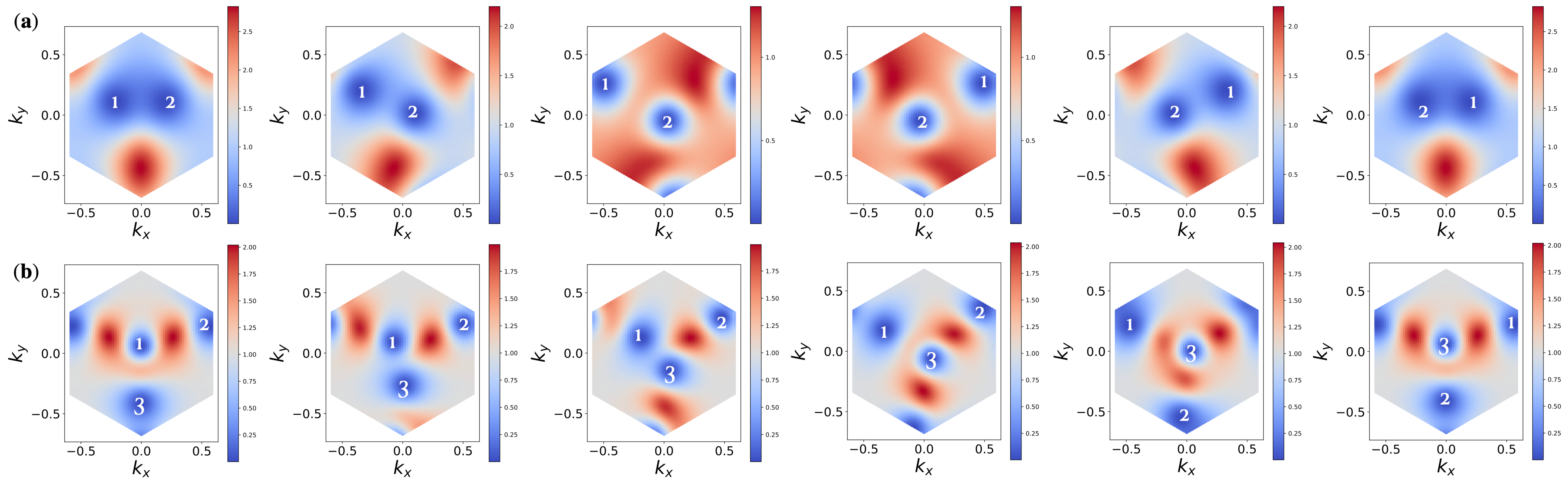}
\caption{Momentum-space zeros and their exchanges. Each figure plots $|u^{\rm Bloch}_{\bm k}(\bm r)|$ as a function of $\bm k$ in the Brillouin zone with position $\bm r$ fixed. For each panel, from left to right, the position $\bm r$ changes from $\bm r_0 = (0.5,0.5)a_M$ to $\bm r_0 + \bm a$ where $a_M$ is the \mr lattice constant and $\bm a$ is a lattice vector. In panel (a), the wavefunction is taken from the chiral twisted bilayer graphene model ($\mC=2$) and $\bm a = -\bm a_1-\bm a_2$. In panel (b), the wavefunction is taken from the chiral twisted trilayer graphene model ($\mC=3$) and $\bm a = \bm a_1$. There are in total $\mC$ zeros in the Brillouin zone at fixed position $\bm r$, and their pattern is lattice translational invariant but indices can be exchanged. We comment that such nodal structure is only visible when the ``unnecessary complications'' from the physical layers of the model is removed by the projection procedure defined in Sec.~\ref{sec:motivation_summary}. The motion of zeros induced by varying position is the dual version of the Thouless pump.}\label{fig_pattern_zero}
\end{figure*}

\subsection{General form of ideal flatband wavefunctions}
Although all $v^{\bm\sigma}_{\bm k}(\bm r)$ satisfies the required momentum-space boundary condition, they violate the real-space lattice translational invariance Eq.~(\ref{def_rbc_uk}). Thereby they individually cannot be the wavefunction for ideal flatbands. In this section, we derive flatband wavefunctions satisfying both the momentum-space and real-space translational properties.

Since at any fixed $\bm r$ basis functions $v^{\bm\sigma}_{\bm k}(\bm r)$ span the vector space of the holomorphic line bundles specified by the momentum-space boundary condition Eq.~(\ref{def_kbc_uk}), wavefunction $u_{\bm k}(\bm r)$ can be definitely written as:
\begin{equation}
    u_{\bm k}(\bm r) = \sum_{\bm\sigma}\mathcal B_{\bm\sigma}(\bm r)v^{\bm\sigma}_{\bm k}(\bm r),
\end{equation}
where $\mathcal B_{\bm\sigma}(\bm r)$ is the linear superposition coefficient that varies as a function of $\bm r$. For the time being we assume $\mathcal B_{\bm\sigma}(\bm r)$ of different colors $\bm\sigma$ are independent functions.

We then impose the real-space boundary condition Eq.~(\ref{def_rbc_uk}). The invariance of $u_{\bm k}(\bm r)$ under lattice translations by $\bm a_{1,2}$ imposes the following constrains:
\begin{eqnarray}
    \mathcal B_{\bm\sigma}(\bm r + \bm a_1) &=& \mathcal B_{\bm\sigma+(1,0)}(\bm r),\label{Bconstrain1}\\
    \mathcal B_{\bm\sigma}(\bm r + \bm a_2) &=& \mathcal B_{\bm\sigma+(0,1)}(\bm r),\label{Bconstrain2}\\
    \mathcal B_{\bm\sigma}(\bm r + \tbm a_i) &=& -e^{-\frac{i}{2\mC}\tbm a_i\times(\bm r+\bm a_{\bm\sigma})}\mathcal B_{\bm\sigma}(\bm r).\label{Bconstrain3}
\end{eqnarray}
These constrain $\mathcal B_{\bm\sigma}(\bm r)$ to have the following form:
\begin{eqnarray}
    \mathcal B_{\bm\sigma}(\bm r) &\equiv& \mathcal B(\bm r+\bm a_{\bm\sigma}),\label{B_defBsigma}\\
    \mB(\bm r+\tbm a_i) &=& -e^{-\frac{i}{2\mC}\tbm a_i\times\bm r}\mB(\bm r),\label{B_quasiperiodic}
\end{eqnarray}
where $\mathcal B(\bm r) = \mathcal B_{\bm\sigma=\bm 0}(\bm r)$. Therefore the wavefunction form is uniquely determined by function $\mathcal B(\bm r)$~\footnote{These functions can be constructed as follows $\mathcal B(\bm r) = g(\bm r)v^*_{\bm k_0}(\bm r)$ where $g(\bm r)$ is $\tbm a-$lattice periodic. The $\bm k_0$ is a fixed, momentum independent tuning parameter. This gives a concrete way to construct the wavefunction and to examine their properties. Indeed the pattern of zeros and their exchange can be numerically verified from wavefunctions constructed in this way.}.

Now we have proved that ideal flatband wavefunctions satisfying the boundary conditions Eq.~(\ref{def_rbc_uk}) and Eq.~(\ref{def_kbc_uk}) has the general form of,
\begin{equation}
    u_{\bm k}(\bm r) = \sum_{\sigma_1=0}^{\mC_1-1}\sum_{\sigma_2=0}^{\mC_2-1}\mB(\bm r+\bm a_{\bm\sigma})v_{\bm k}(\bm r + \bm a_{\bm\sigma}),\label{ansatzuk}
\end{equation}
where $\mathcal B(\bm r)$ is a $\bm k-$independent quasi-periodic function satisfying Eq.~(\ref{B_quasiperiodic}), and it tunes the Berry curvature. It is now obvious that factorization $\mC_{1,2}$ is a gauge choice: difference choice of factorization affects basis $v^{\bm\sigma}_{\bm k}$ whereas $u_{\bm k}$ is basis independent. When taking $\mC=1$, the wavefunction reduces to the form consistent with the result derived in Ref.~\cite{JieWang_exactlldescription}. Our wavefunction Eq.~(\ref{ansatzuk}) is in the explicit first quantized form and is a generalization of the previously derived ``color-entangled wavefunction'' with constant Berry curvature~\cite{Yangle_Modelwf,Yangle_haldanestatistics,Maissam_Qi_PRX12}. Here ``color'' refers to the LLL wavefunctions $v^{\bm\sigma}_{\bm k}$ and translations ``entangles'' them.

Now we comment on the pattern of zeros and their exchanges observed from the wavefunction of the chiral twisted multilayer graphene model shown in Fig.~\ref{fig_pattern_zero}: since lattice translation can permute the color index, zeros are rearranged under one period of such lattice translation; the flatband wavefunction $u_{\bm k}$, which is a summation of all colors components, remains invariant so the pattern of all zeros is lattice translation periodic.

We can label the location of the $\mC-$zeros in the complex momentum-space by $\zeta_{i=1,...,\mC}(\bm r)$ which are $\mC$ smooth functions maps coordinate $\bm r$ to $k=\omega^a\bm k_a$. There is an important consistency check we can perform. As a consequence of the Abel theorem, the sum of the positions of the zeros of each of our basis wavefunctions shall add up to the boundary condition modulo the lattice~\footnote{This can be derived based on the explicit wavefunction form and by using the momentum-space boundary condition Eq.~(\ref{def_kbc_uk}).},
\begin{equation}
    -i\sum_{i=1}^{\mC} \zeta_{i}(\bm r) = z \mod ~ m_1 a_1 + m_2 a_2.
\end{equation}
We anticipate the zeros $\zeta_i(\bm r)$ are useful in constructing explicit first-quantized Halperin type wavefunction by regarding index $i$ as layers. We leave this for future exploration.

To conclude this section, relying on the emergent dual guiding centers and their simple algebra, we derived the most general form of Chern number $\mC$ ideal flatband wavefunction in Eq.~(\ref{ansatzuk}). It is given by $\mC$ LLL type wavefunctions entangled by lattice translations. During such derivation, we pointed out the importance of adopting a position-momentum exchanged view for flatband.

\section{Interacting physics: density operator, exact GMP algebra and model fractional Chern insulators\label{sec:densityalgebra}}
Motivated by the unusual numerical observation discussed in Sec.~\ref{sec:motivation_summary}, in this section we carefully examine the projected density operator and their algebra. Utilizing the universal form of the ideal flatband wavefunction derived in the previous section, here we explicitly derive the projected density operator and their algebra for all topological ideal flatbands. Importantly, we show their density operator obeys a closed algebra, generalizing the GMP algebra in a nontrivial way. The generalized GMP algebra is the fundamental reason in giving rise to the model FCIs. This result strongly disproved the common lore that fluctuating quantum geometries destabilize FCI type many-body phases of matter.

\subsection{Definition of the density operator}
We define the $\tmH$-space projected density operator as follows:
\begin{equation}
    \hat\rho_{\bm q} = \sum_{\bm k}\langle u_{\bm k+\bm q}|u_{\bm k}\rangle c^{\dag}_{\bm k+\bm q}c_{\bm k},\label{def_projrho}
\end{equation}
and denote the flatband projected density operator as,
\begin{equation}
    \hat\rho^{\rm Bloch}_{\bm q} = \sum_{\bm k}\langle u^{\rm Bloch}_{\bm k+\bm q}|u^{\rm Bloch}_{\bm k}\rangle c^{\dag}_{\bm k+\bm q}c_{\bm k}.\label{def_projrho_Bloch}
\end{equation}
Here $u_{\bm k}\in\tmH$ and $u^{\rm Bloch}_{\bm k}\in\mathcal H$ are defined in Eq.~(\ref{defuk}). These two density operators differ by $\bm k-$dependent normalization factors in their matrix element. We will focus on $\hat\rho_{\bm q}$ and will justify why normalization factor is unimportant for the many-body zero modes.

The $\hat c^\dag_{\bm k}$ and $\hat c_{\bm k}$ are the standard particle creation, annihilation operator obeying the standard commutation relation: for fermions we have $\{\hat c_{\bm k}, \hat c_{\bm k'}\} = 0$ and for bosons $[\hat c_{\bm k}, \hat c_{\bm k'}] = 0$. Since they create/ annihilate the wavefunction $u_{\bm k}$, they must obey a boundary condition inherited from Eq.~(\ref{def_kbc_uk}):
\begin{equation}
    c^\dag_{\bm k+\bm b} = \eta_{\bm b} e^{\frac{i\mC}{2}\bm b\times\bm k} c^\dag_{\bm k},\label{def_cdag_kbc}
\end{equation}
where $\eta_{\tbm b}$ is plus or minus one valued: for a vector on the $\tbm b-$lattice (which of course include the $\bm b-$lattice),
\begin{equation}
    \eta_{\tbm b} = (-1)^{m+n+mn} \quad \text{for} \quad \tbm b = m\tbm b_1+n\tbm b_2.\label{def_eta}
\end{equation}

\subsection{Emergent exact GMP algebra at $\mC=1$\label{sec:sub:exactgmpc1}}
We begin with $\mC=1$ ideal flatbands as their color space is simple with unit dimension. We then generalize to generic $\mC>1$ ideal flatbands. We can drop the color index in this case. Reducing Eq.~(\ref{ansatzuk}) to $\mC=1$ we get its wavefunction previously derived in Ref.~\cite{JieWang_exactlldescription}:
\begin{equation}
    u_{\bm k}(\bm r) = \mathcal B(\bm r)\Phi_{\bm k}(\bm r),\label{conewf}
\end{equation}
where $\Phi_{\bm k}(\bm r)$ is the LLL wavefunction of magnetic length $l_B=1$. Since for $\mC=1$ there is no difference between the $\bm a-$lattice and the $\tbm a-$lattice, implied from Eq.~(\ref{B_quasiperiodic}) the $\mathcal B(\bm r)$ is quasi-periodic function satisfying,
\begin{equation}
    \mathcal B(\bm r + \bm a_i) = -e^{-\frac{i}{2}\bm a_i\times\bm r}\mathcal B(\bm r).
\end{equation}

We define a lattice periodic function which admits a Fourier transform as follows,
\begin{equation}
    \mathcal A(\bm r) \equiv |\mathcal B(\bm r)|^2 = \sum_{\bm b}\omega_{\bm b}e^{i\bm b\cdot\bm r},\label{cone_fourier}
\end{equation}
where $\omega_{\bm b}$ are the unique tuning parameters in the $\mC=1$ problem. Eq.~(\ref{conewf}) implies the flatband form factor can be expressed in terms of LLL form factors:
\begin{equation}
    \langle u_{\bm k}|u_{\bm k'}\rangle = \sum_{\bm b} \omega_{\bm b} \langle\Phi_{\bm k}|e^{i\bm b\cdot\hat{\bm r}}|\Phi_{\bm k'}\rangle = \sum_{\bm b} \omega_{\bm b} f^{\bm k\bm k'}_{-\bm b},
\end{equation}
where the form factor of the LLL wavefunction is given in below,
\begin{eqnarray}
    f^{\bm k\bm k'}_{-\bm b} &\equiv& \int d^2\bm r~e^{i\bm b\cdot\bm r}\Phi^*_{\bm k}(\bm r)\Phi_{\bm k'}(\bm r),\label{def_f}\\
    &=& \eta_{\bm b}e^{-\frac{i}{2}(\bm k+\bm k')\times\bm bl_B^2}e^{\frac{i}{2}\bm k\times\bm k'l_B^2}e^{-\frac{1}{4}|\bm k-\bm k'+\bm b|^2l_B^2}.\nonumber
\end{eqnarray}

With this form factor, we can rewrite the flatband density operator as follows:
\begin{equation}
    \hat\rho_{\bm q} = \sum_{\bm b} \omega_{\bm b}\hat\rho^{\rm GMP}_{\bm q + \bm b},\label{cone_gmp}
\end{equation}
where $\hat\rho^{\rm GMP}_{\bm q} = \sum_{\bm k} f^{\bm k+\bm q,\bm k}_{\bm 0}c^\dag_{\bm k+\bm q}c_{\bm k}$ is the LLL density operator obeying the GMP algebra Eq.~(\ref{projectedGMP}). In deriving Eq.~(\ref{cone_gmp}), the boundary condition Eq.~(\ref{def_cdag_kbc}) is used.

Eq.~(\ref{cone_gmp}) is one of the key result of this section: it shows that the flatband density operator $\hat\rho_{\bm q}$ is obtained from downfolding the LLL density operator into the Brillouin zone and the downfolding coefficients $\omega_{\bm b}$ are the tuning parameters for the Berry curvature. It also provides an explanation to the exact FQH type FCIs occurring in the $\mC=1$ ideal flatbands, alternative to the approach from constructing many-body wavefunctions~\cite{Grisha_TBG2}. To see this, we expand the flatband density-density interaction as follows:
\begin{eqnarray}
    H &=& \sum_{\bm b} |\omega_{\bm b}|^2 \cdot \hat\rho^{\rm GMP}_{\bm q + \bm b} \hat\rho^{\rm GMP}_{-\bm q - \bm b} + \sum_{\bm b\neq\bm 0} \omega^2_{\bm b} \cdot \hat\rho^{\rm GMP}_{\bm q + \bm b} \hat\rho^{\rm GMP}_{-\bm q + \bm b},\nonumber\\
    &+& \sum_{\bm b\neq\pm\bm b'} \omega_{\bm b} \omega^*_{\bm b'} \cdot \hat\rho^{\rm GMP}_{\bm q + \bm b} \hat\rho^{\rm GMP}_{-\bm q - \bm b'}.\label{Hconepp}
\end{eqnarray}

The two terms in the first line are the modified relative interaction and the center-of-mass (COM) interaction~\cite{JieWang_exactlldescription}, respectively.

The first term scatters particles by relative momentum $\bm q+\bm b$ while preserving their COM; the strength of interaction is modified by the coefficient $\omega_{\bm b}$ but the power (range) of the interaction remains unchanged. Therefore this part modify the zero mode.

The second term represents a new type of interaction: the two-particle interaction depends not only on their relative distance $\hat{\bm R}_- \equiv \hat{\bm R}_1 - \hat{\bm R}_2$ but also on their COM $\hat{\bm R}_+ \equiv \hat{\bm R}_1 + \hat{\bm R}_2$. Since the relative guiding centers commutes with the COM, two sets of independent pseudopotential operators can be constructed,
\begin{eqnarray}
    \hat P^+_M &=& 2\int \frac{d^2\bm ql_B^2}{(2\pi)^2} L_M(\bm q^2)e^{-\frac12\bm ql_B^2}e^{i\bm q\cdot\hat{\bm R}_+},\\
    \hat P^-_m &=& 2\int \frac{d^2\bm ql_B^2}{(2\pi)^2} L_m(\bm q^2)e^{-\frac12\bm ql_B^2}e^{i\bm q\cdot\hat{\bm R}_-},
\end{eqnarray}
where $L_n(x)$ is the $n_{\rm th}$ Laguerre polynomial. Integration over $\bm q$ is replaced with lattice sum for finite systems. The exact GMP algebra ensures that $P^{\pm}$ individually are exact projector that projects two particles into their COM angular momentum $M$ and relative momentum $m$ channel, respectively. Since the FQH ground states are exactly annihilated by the relative projectors $\hat P^-_m$ and insensitive to the COM projector, the inclusion of such COM interaction also cannot modify the existence of zero modes. The last line of Eq.~(\ref{Hconepp}) is a mixtures of these two contributions, which also does not affect the interacting zero modes. More detailed discussion on the COM interaction and the generalized FQH model can be found in Ref.~\cite{JieWang_exactlldescription}.

Last but not least, we justify the statement that ignoring the normalization factor cannot not affect the dimension of interacting zero modes. For unnormalized basis, or generally speaking for non-orthonormal basis, the eigen-equation is described by the following:
\begin{equation}
    \sum_{\beta} H_{\alpha\beta}c_{\beta} = \lambda \sum_{\beta}O_{\alpha\beta}c_{\beta},\label{gen_eigen}
\end{equation}
where $\alpha, \beta$ denote the entry of the Hamiltonian, $O$ is the basis overlap matrix and $\lambda$ is the eigenvalue. As long as $O$ is full ranked, the $\lambda=0$ eigenvalue, if exist, is unaffected by the concrete form of $O$. Therefore we arrived a useful message for our problem: treating an non-orthonormal system as orthonormal does not affect the dimension of zero modes.



\subsection{Emergent exact GMP algebra at general $\mC\geq1$ and family of model Hamiltonians}
The physics of higher Chern bands is enriched by their color degrees of freedom. We start with the discussion about the tuning parameter for the Berry curvature, followed by discussing the general $\mC\neq1$ density operator and their algebras.

\subsubsection{Tuning parameter for Berry curvature}
To extract the useful information for Berry curvature, we define,
\begin{equation}
    \mathcal A_{\bm\sigma\bm\sigma'}(\bm r) \equiv e^{\frac{i\mC}{2}\bm q_{\bm\sigma}\times\bm q_{\bm\sigma'}}\mathcal B^*_{\bm\sigma}(\bm r)\mathcal B_{\bm\sigma'}(\bm r)e^{-\frac{i\mC}{2}\left(\bm q_{\bm\sigma}-\bm q_{\bm\sigma'}\right)\cdot\bm r},
\end{equation}
where
\begin{equation}
    \left(\bm q_{\bm\sigma}\right)_a = \mC \epsilon_{ab}\left(\bm a_{\bm\sigma}\right)^b = \left(-\frac{\sigma_1}{\mC_1}\bm b_2 + \frac{\sigma_2}{\mC_2}\bm b_1\right)_a.
\end{equation}
Based on the quasi-periodicity Eq.~(\ref{B_quasiperiodic}), it is easy to show that $\mathcal A$ is $\tbm a-$lattice translational symmetric satisfying $\mathcal A_{\bm\sigma\bm\sigma'}(\bm r) = \mathcal A_{\bm\sigma\bm\sigma'}(\bm r+\tbm a)$. Here $\tbm a$ denotes a generic lattice vector $\tbm a = m\tbm a_1 + n\tbm a_2$. As a consequence, $\mathcal A_{\bm\sigma\bm\sigma'}$ can be decomposed into Fourier modes on the $\tbm b-$lattice:
\begin{equation}
    \mathcal A_{\bm\sigma\bm\sigma'}(\bm r) = \sum_{\tbm b}\omega_{\tbm b}(\bm\sigma,\bm\sigma')e^{i\tbm b\cdot\bm r}.\label{defAfourier}
\end{equation}
Compared with Eq.~(\ref{cone_fourier}), the Fourier modes here are enriched by the color indices and are defined on a smaller momentum-space lattice scale. We remind the reader this is because for $\mC>1$ the LLL wavefunctions $v^{\bm\sigma}_{\bm k}$ has an enlarged magnetic length $l_B^2 = \mC$. See Fig.~\ref{fig_BZ} (c) for an illustration of the correspondence between flatband and LLL length scales.

In fact, not all coefficients $\omega_{\tbm b}(\bm\sigma,\bm\sigma')$ are independent because colors can be related by lattice translations. It turns out after straightforward Fourier transformation analysis that, the independent tuning parameters of the system are given by $\omega_{\tbm b}(\bm 0,\delta\bm\sigma)$ where $\delta\bm\sigma \equiv \bm\sigma'-\bm\sigma$ is the color difference taking values from $\delta\sigma_i\in[0,\mC_i-1]$. All other components can be generated via:
\begin{equation}
    \omega_{\tbm b}(\bm\sigma,\bm\sigma+\delta\bm\sigma) = \omega_{\tbm b}(\bm 0,\delta\bm\sigma)\times e^{i\mC\bm q_{\bm\sigma}\times\tbm b}.\label{sym_omega1}
\end{equation}
We therefore can focus on the $\omega_{\tbm b}(\bm 0, \delta\bm\sigma)$ components. Soon we will find it is particularly convenient to define and use the new notation:
\begin{equation}
    \omega_{\tbm b}[\bm q_{\delta\bm\sigma}] \equiv \omega_{\tbm b}(\bm 0, \delta\bm\sigma).
\end{equation}

The color space is periodic as the functions $\mathcal B_{\bm\sigma}$ are all quasi-periodic when $\bm r$ advanced by a lattice vector $\tbm a$. This puts another useful relation to the Fourier components that allows us to move from the fractional lattice $\tbm b$ to integer lattice $\bm b$. Such relation is:
\begin{equation}
    \omega_{\bm b+\tbm b'}[\bm q_{\delta\bm\sigma}] = \eta_{\tbm b'}e^{-\frac{i\mC}{2}\tbm b'\times\bm q_{\delta\bm\sigma}}\times\omega_{\bm b}[\bm q_{\delta\bm\sigma}-\tbm b'].\label{sym_omega2}
\end{equation}
We leave detailed derivations of Eq.~(\ref{sym_omega1}) and Eq.~(\ref{sym_omega2}) into to the appendix.

An important implication from Eq.~(\ref{sym_omega2}) is that, for any lattice vectors $\tbm b$, one can always factorize it into an ``integer part'' $\bm b$ living on the $\bm b-$lattice and a remaining ``fractional part'' $\tbm b'$ defined within $\bm b_{1,2}$, such that $\tbm b = \bm b + \tbm b'$. Then Eq.~(\ref{sym_omega2}) says the fractional lattice dependence can be effectively ``absorbed'' into the color space. To conclude, we find the $\mC>1$ system is enriched by the color degrees of freedom, and useful information about band geometries are contained either in the set of parameters:
\begin{equation}
    \{\omega_{\tbm b}(\bm 0,\delta\bm\sigma)~|~\tbm b = m\tbm b_1+n\tbm b_2;~\delta\sigma_i\in[0,\mC_i-1]\},
\end{equation}
or equivalently in the set of parameters:
\begin{equation}
    \{\omega_{\bm b}[\bm q_{\delta\bm\sigma}]~|~\bm b = m\bm b_1+n\bm b_2;~\delta\sigma_i\in[0,\mC-1]\}.
\end{equation}

We will find $\{\omega_{\tbm b}(\bm 0,\delta\bm\sigma)\}$ convenient for numerical calculation as the color space is smaller, and $\{\omega_{\bm b}[\bm q_{\delta\bm\sigma}]\}$ convenient in analytically deriving the density algebra because of the absence of fractional lattice $\tbm b-$dependence.

\subsubsection{Family of exact parent Hamiltonians}
We proceed to discuss interacting physics in $\mC>1$ ideal flatbands. The explicit form of the wavefunction and wavefunction overlap enable us to derive the explicit form of the flatband projected interacting Hamiltonian. In this section, we focus on the two-body density-density interaction Eq.~(\ref{defH}) but the methods here applies to generic $M-$body interactions. We derive a family of Hamiltonians, whose Berry curvature is tunable, while all of them preserving model FCIs as the exact zero-energy eigenstates for short-ranged interactions. Therefore, these Hamiltonians are exact parent Hamiltonians for model FCIs. In the following section, we derive the density algebra for general $\mC>1$ and provides a theoretical understanding for the origin of these model FCIs.

The flatband projected density operator is defined in Eq.~(\ref{def_projrho}). For numerical calculations, all degrees of freedom must be restricted into the $N_1N_2$ orthogonal states defined within the first Brillouin zone of the flatband. Thereby we need to use the boundary condition Eq.~(\ref{def_kbc_uk}) to convert the momentum of $c^\dag_{\bm k}$ appearing in Eq.~(\ref{def_projrho}) into the first Brillouin zone. The resulting density-density interacting Hamiltonian for numerical diagonalization study is,
\begin{equation}
    H = \sum_{\bm b}\sum'_{\bm k_1,...,\bm k_4}v_{\bm k_1-\bm k_4-\bm b}\mF^{\bm k_1\bm k_4}_{\bm b}\mF^{\bm k_2\bm k_3}_{-\bm b+\delta\bm b} c^\dag_{\bm k_1}c^\dag_{\bm k_2}c_{\bm k_3}c_{\bm k_4},\label{def_interactingH}
\end{equation}
where $\delta\bm b=\bm k_1+\bm k_2-\bm k_3-\bm k_4$. The $\sum'_{\bm k}$ sums momentum $\bm k$ defined in the first Brillouin zone, $\sum_{\bm b}$ sums reciprocal lattice vectors $\bm b$ of the entire 2D momentum space. In the above, $\mF$ is defined and expressed as:
\begin{eqnarray}
    \mF^{\bm k\bm k'}_{\bm b} &=& \int d^2\bm r~e^{-i\bm b\cdot\bm r}u^*_{\bm k}(\bm r)u_{\bm k'}(\bm r),\\
    &=& \sum_{\bm\sigma\bm\sigma'}\sum_{\tbm b}\omega_{\tbm b}(\bm\sigma,\bm\sigma')g^{\bm k\bm k'}_{\bm b-\tbm b}(\bm\sigma,\bm\sigma'),\nonumber
\end{eqnarray}
and the function $g$ is:
\begin{equation}
    g^{\bm k\bm k'}_{\tbm b}(\bm\sigma,\bm\sigma') = e^{\frac{i\mC}{2}\left(-\bm q_{\bm\sigma}\times\bm q_{\bm\sigma'} + \bm q_{\bm\sigma}\times\bm k - \bm q_{\bm\sigma'}\times\bm k'\right)}f^{\bm k+\bm q_{\bm\sigma},\bm k'+\bm q_{\bm\sigma'}}_{\tbm b},\label{defg}
\end{equation}
where $f$ is the LLL form factor given in Eq.~(\ref{def_f}) with $l_B^2 = \mC$ and $\eta_{\tbm b}$ is defined in Eq.~(\ref{def_eta}). The $\omega_{\tbm b}(\bm\sigma,\bm\sigma')$ are the tuning parameters of the model discussed in the last section. They are equivalently parameterized by $\omega_{\bm b}[\bm q_{\delta\bm\sigma}]$.

Our interacting Hamiltonian Eq.~(\ref{def_interactingH}) is a generalization of the previous studied color-entangled Hamiltonian~\cite{Yangle_Modelwf,Yangle_haldanestatistics} to allow nonuniform Berry curvature. The model of Ref.~\cite{Yangle_Modelwf} has constant Berry curvature and corresponds to set $\omega_{\bm b}[\bm q_{\delta\bm\sigma}] = \delta_{\bm b,\bm 0}\delta_{\delta\bm\sigma,\bm 0}$ in our model.

We have numerically verified that for all parameters $\omega_{\bm b}[\bm q_{\delta\bm\sigma}]$, our model Hamiltonian Eq.~(\ref{def_interactingH}) exhibits exact $\nu^{-1} = [(m+1)\mC + 1]$ fold zero-energy degeneracy at filling fraction $\nu$ for the short-ranged $v_m$ interaction. Particle-cut entanglement spectra analysis shows that these exact FCIs are Halperin-type states. Moreover, we numerically find multi-body short-ranged interactions support non-Abelian model FCIs, such as analogs of Read-Rezayi series~\cite{Read_Rezayi} and non-Abelian spin singlet state~\cite{Ardonne_Schoutens_NASS}, as exact zero-energy eigenstates for arbitrary $\omega_{\bm b}[\bm q_{\delta\bm\sigma}]$, suggesting the momentum-space complex structure is the origin of exact FCIs for generic $M$-body repulsions and should be a new fundamental property of density operators. In the next sections, we analytically derive the density operator and discuss their hidden exact GMP algebra.

\subsubsection{Density operator and closed algebra}
The general form of Chern $\mC$ ideal flatband wavefunction is a superposition of $\mC$ LLL type wavefunctions. Using this fact and the Fourier modes $\omega_{\bm b}[\bm q_{\bm\sigma}]$ defined in the previous section, the density operator can be directly computed. After some algebra detailed in the appendix, we find the flatband density operator is given by as follows:
%
%
\begin{equation}
    \hat\rho_{\bm q} = \sum'_{d\bm\sigma}\sum_{\bm b}\omega_{\bm b}[\bm q_{\delta\bm\sigma}]~\hat\rho^{\rm GMP}_{\bm q+\bm b}(\delta\bm\sigma),\label{rho1}
\end{equation}
where $\sum'$ sums color from $\delta\sigma_i\in[0,\mC-1]$. It is downfolded from $\hat\rho^{\rm GMP}$ into the flatband Brillouin zone, and the downfolding coefficient $\omega_{\bm b}[\bm q_{\bm\sigma}]$ is precisely the coefficient that controls the Berry curvature distribution in the flatband problem. The $\hat\rho^{\rm GMP}$ is can be expressed as,
\begin{equation}
    \hat\rho^{\rm GMP}_{\bm q}(\delta\bm\sigma) = \sum_{\bm k}g^{\bm k+\bm q,\bm k}(\delta\bm\sigma)~c^\dag_{\bm k+\bm q}c_{\bm k},
\end{equation}
where $g^{\bm k\bm k'}(\delta\bm\sigma) \equiv g^{\bm k\bm k'}_{\bm 0}(\bm0,\delta\bm\sigma)$ defined in Eq.~(\ref{defg}). The physical meaning of the density operator will be clarified in the next section.

Importantly $\hat\rho^{\rm GMP}$ obeys a closed algebra of Girvin-MacDonald-Platzman type. Denoting $q \equiv \omega^a\bm q_a$ as the complex variable, we have,
\begin{widetext}
    \begin{equation}
    [\hat\rho^{\rm GMP}_{\bm q_1}(\delta\bm\sigma_1), \hat\rho^{\rm GMP}_{\bm q_2}(\delta\bm\sigma_2)] = \left[e^{-\frac{i\mC}{2}\bm q_{\delta\bm\sigma_1}\times\bm q_{\delta\bm\sigma_2}}e^{\mC\left(q_1-q_{\delta\sigma_1}\right)^*\left(q_2-q_{\delta\sigma_2}\right)}-h.c.\right]\times\hat\rho^{\rm GMP}_{\bm q_1+\bm q_2}(\delta\bm\sigma_1+\delta\bm\sigma_2).\label{rho2}
    \end{equation}
\end{widetext}

Eq.~(\ref{rho1}) and Eq.~(\ref{rho2}) are key results of this work. They are valid for any system size $N_{1,2}$ and Chern number factorization $\mC_{1,2}\geq 1$. They follow directly from the ideal quantum geometric condition Eq.~(\ref{defidealcond}) without using any further assumptions~\footnote{Generalizing to negative Chern number is straightforward as single-particle wavefunction is simply complex conjugated.}. They show how Berry curvature fluctuation (controlled by $\omega_{\bm b}[\bm q_{\bm\sigma}]$) influences the projected density operator and how the exact GMP algebra emergence in the Hilbert space $\tmH$ of ideal flatbands. 

\subsubsection{Mapping to multilayer Landau levels}
In this section, we discuss the physical interpretation of density operator $\hat\rho^{\rm GMP}$, and discuss how the GMP algebra gives rise to exact FCIs. We first recall that the colors, {\it i.e.} the basis functions, $v^{\bm\sigma}_{\bm k}$ are related to each other by lattice translations, and each of them has a magnetic unit cell $\tbm a_{1,2}$ which encloses an area $\mC$ times larger than the lattice unit cell. Thereby their magnetic Brillouin zone is $\mC$ times smaller than the flatband Brillouin zone. For what follows we assume commensurate geometry such that $N_i$ is divisible by $\mC_i$. Incommensurate geometries can be easily turned into commensurate geometries by gluing multiple systems together. Such gluing merely changes the spatial periodicity of the interaction, which we argue will not affect interacting zero modes as they are only sensitive to the short-ranged component of interaction. See Fig.~\ref{fig_BZ} (c) and (d) for an illustration of $\mC=2$.

The colors are LLL of $\mC$ distinct boundary conditions. To see this, we use them to define $\Phi^{\bm\sigma}_{\bm k}$:
\begin{equation}
\Phi_{\bm k}^{\bm\sigma}(\bm r) \equiv e^{i\bm k\cdot\bm r}e^{-\frac{i}{2\mC}\bm a_{\bm\sigma}\times\bm r}v_{\bm k}(\bm r+\bm a_{\bm\sigma}),
\end{equation}
which can be equivalently expressed by:
\begin{equation}
    \Phi_{\bm k}^{\bm\sigma}(\bm r) = e^{-\frac{i\mC}{2}\bm q_{\bm\sigma}\times\bm k}\Phi_{\bm k+\bm q_{\bm\sigma}}(\bm r).\label{def_Phiksigma}
\end{equation}
They satisfy the following minimal and maximal boundary conditions:
\begin{eqnarray}
    e^{i\tbm b_i\cdot\bm R}|\Phi_{\bm k}^{\bm\sigma}\rangle &=& -e^{-2\pi i\sigma_i/\mC_i}\cdot e^{i\tbm b_i\times\bm k}|\Phi_{\bm k}^{\bm\sigma}\rangle,\label{defPhiksigma1}\\
    e^{i\bm b_i\cdot\bm R}|\Phi_{\bm k}^{\bm\sigma}\rangle &=& (-1)^{\mC_i}|\Phi_{\bm k}^{\bm\sigma}\rangle.\label{defPhiksigma2}
\end{eqnarray}
This means: given maximal boundary condition Eq.~(\ref{defPhiksigma2}), colors are distinguished by their $\mC$ different minimal boundary conditions thus living in $\mC$ different Hilbert spaces $\mathcal H_{\bm\sigma}$ with $\sigma_i\in[0,\mC_i-1]$. Such $\mC$ spaces $\mathcal H_{\bm\sigma}$ are not necessarily orthogonal but are independent, thereby fully spanning the flatband Hilbert space $\tmH$ and is an equivalent description of flatband physics. Therefore when $\mC>1$, we are dealing with a problem with multiple boundary conditions. Then we come to a subtly. Remember the notion of momentum is itself a gauge choice {\it i.e.} it is boundary condition dependent. This is clearly seen from Eq.~(\ref{def_Phiksigma}): although $v^{\bm\sigma}_{\bm k}$ or $\Phi^{\bm\sigma}_{\bm k}$ is indexed by a momentum $\bm k$, their momentum measured from the $\mathcal H_{\bm\sigma=\bm 0}$ space is in fact $\bm k+\bm q_{\bm\sigma}$. To make statements clear, we call the momentum measured from the $\bm\sigma=\bm 0$ frame as the ``absolute momentum'' and momentum measured from individual $\mathcal H_{\bm\sigma}$ as ``relative momemtum''. So $\Phi^{\bm\sigma}_{\bm k}$ has relative momentum $\bm k$ and absolute momentum $\bm k+\bm q_{\bm\sigma}$. 

The density operator $\hat\rho^{\rm GMP}_{\bm q}(\delta\bm\sigma)$ then has a simple interpretation: it boosts the absolute momentum of a particle in $\mathcal H_{-\bm\sigma'}$ by $\bm q$ and at the same time maps it into $\mathcal H_{-\bm\sigma=-\bm\sigma'+\delta\bm\sigma}$. To see this explicitly, we can rewrite the density operator as,
\begin{eqnarray}
    &&\hat\rho^{\rm GMP}_{\bm q}(\bm\sigma'-\bm\sigma),\nonumber\\
    &=& \sum_{\bm k}g^{\bm k+\bm q,\bm k}(\bm\sigma'-\bm\sigma)|\bm k+\bm q\rangle\langle\bm k|,\\
    &=& e^{-\frac{i\mC}{2}\bm q_{\bm\sigma}\times\bm q_{\bm\sigma'}} \sum_{\bm k} f^{\bm k+\bm q+\bm q_{\bm\sigma}, \bm k+\bm q_{\bm\sigma'}}\cdot|\Phi^{-\bm\sigma}_{\bm k+\bm q+\bm q_{\bm\sigma}}\rangle\langle\Phi^{-\bm\sigma'}_{\bm k+\bm q_{\bm\sigma'}}|,\nonumber
\end{eqnarray}
where $|\bm k\rangle\in\tmH$ is a flatband state and $|\Phi^{-\bm\sigma}_{\bm k}\rangle\in\mathcal H_{-\bm\sigma}$ is a LLL states, and $f$ is the LLL form factor. We thus justified the meaning of the density operator discussed above.

The resulting interacting Hamiltonian in the LL basis is however not standard, but interacting zero modes can still be understood. Firstly, the downfolding induced by Berry curvature Eq.~(\ref{rho1}) implies there is a center-of-mass interaction in this generalized FQH problem. Such center-of-mass interaction cannot affect the zero mode for the same reason as explained in the $\mC=1$ case discussed in Sec.~\ref{sec:sub:exactgmpc1} and Ref.~\cite{JieWang_exactlldescription}. Secondly, different from the layers in LL problems, here colors are not orthogonal. The non-orthogonality also cannot affect zero modes; see discussions around Eq.~(\ref{gen_eigen}). Lastly, the density operators in multilayer LL problems are diagonal in layer, whereas in flatbands the density operator is not diagonal in colors. We have numerically verified this also does not affect the dimension of zero modes.


To conclude this section, using the explicit form of the universal color-entangled wavefunction proved in Sec.~\ref{sec:idealwavefunction}, in this section we explicitly computed the ideal flatband density operator. We found generally the flatband density is obtained from downfolding $\hat\rho^{\rm GMP}$ into the flatband Brillouin zone in an exact manner. Importantly, the algebra of $\hat\rho^{\rm GMP}$ is closed, generalizing the previously derived GMP algebra from $\mC=1$ LL to the much more general case for ideal flatbands of arbitrary $\mC>0$ in a nontrivial way. We demonstrate the generalized GMP algebra enables an exact mapping from the flatband problem to multilayered LL problem, where color plays the role of layer in such mapping. While the resulting multilayered LL problem is still nonstandard in many aspects (exhibiting center-of-mass dependent interactions, non-orthogonal layer degrees of freedom and non-diagonal density matrix), we show neither of them can modify the dimension of many-body zero modes. This summary also answers the two questions posed in the end of Sec.~\ref{sec:motivation_summary} which motivate the theory presented in this work.

\section{Discussions\label{sec:conclusion}}
The flatband systems are ideal venues to explore strongly interacting phenomena. However the physics is often complicated by the interactions and by the wavefunction's intrinsic geometry such as Berry curvature. Through systematically study flatbands in the ideal limit, we derive a general form of high Chern number Bloch wavefunction that will be useful for describing various types of quantum phase of matter: on one side our wavefunction is inherited from LLL wavefunction thereby it can naturally express FQH type fractionalized states; on the other side the internal color degree of freedom allows them to equally well describe symmetry breaking phases. Moreover we derive a family of exact interacting parent Hamiltonians and point out the hidden closed density algebra. This will provide a general framework and paving the way to thoroughly explore the interplay between wavefunction's geometry and interaction. Our theory also has practical implications to realistic \mr materials and beyond.

Firstly, our results are useful in guiding the experimental search of FCIs in \mr materials. Recently FCIs were reported in twisted bilayer graphene with weak external fields~\cite{FCI_TBG_exp}. The role of ideal quantum geometry in this experiment has been highlighted, mainly motivated by the exact results for the $\mC=1$ ideal flatbands in the twisted bilayer graphene at the chiral limit~\cite{Grisha_TBG,Grisha_TBG2,JieWang_exactlldescription} and numerical results in the realistic parameter regime~\cite{Dan_parker21}. The importance of the ideal geometry for general $\mC>1$ however has not been rigorously justified. Our exact results presented in this work set ideal quantum geometry as one of the most crucial indicators~\cite{Dan_parker21} for experimental realization of generic FCIs in \mr materials~\cite{ZhaoLiu_TBG,ZhaoLiu_TDBG,Cecile_PRL20,Cecile_TBG_Flatband,Kaisun_FCI21,Valentin22_anomaloushallmetal,Lanchli21,Claassen:2022vt}. The key message from our results is that: optimizing the single-particle band width and band geometry to approach the ideal condition Eq.~(\ref{defidealcond}) should be given high priority before employing heavy numerical computation involving interactions.

Secondly, recently the momentum-space geometry is found important not only for FCIs, but also for other phases such as symmetry breaking states~\cite{Young_partiallyfilledTBG}, superconductivity~\cite{Peotta:2015aa,torma_bernevig22,bernevig_torma22,Hofmann:2022wq} and others~\cite{cTBG_stern,Abouelkomsan_ph22,Ewelina22,Bruno_deg22,Bruno_experiment22}. For example, in twisted monolayer/ bilayer graphene systems, various topological and non-topological symmetry breaking states are observed experimentally~\cite{Young_partiallyfilledTBG,FCI_TBG_exp}. In particular, half filling a $\mC=2$ band is experimentally found to have spontaneous translational symmetry breaking to give rise to topological charge density wave. Such topological charge density wave state can be naturally understood from our theory as spontaneous color polarization: as the ideal flatband wavefunction is a nonlinear superposition of $\mC$ LLL states shifted by lattice translations and thus distinguished by their boundary conditions, the real-space charge density wave pattern is a consequence of the polarization in the color space. In general, our single-particle wavefunction can be regarded as the parent wavefunction for various daughter states: generic flatband wavefunction can be argued to be perturbed from this wavefunction by breaking the K\"ahler condition, adding finite dispersion, breaking internal color space symmetries and others. The ideal flatband wavefunction is useful in understanding various symmetry breaking mechanism. Apart from quantum Hall related physics, exploring the implication from ideal quantum geometry for flatband superconductivity~\cite{Peotta:2015aa,FangXie_FlatbandSC,torma_bernevig22,bernevig_torma22,Hofmann:2022wq,danmao22} in double layer time-reversal symmetric high Chern bands is an interesting future topic.

Besides the wavefunction, our derived family of interacting Hamiltonian with tunable Berry curvature also deserves future explorations. They are model Hamiltonians with controlled properties: for arbitrary tuning parameters of the Berry curvature, FCIs are exact zero energy as long as interaction is short-ranged. These model Hamiltonians are thus ideal platforms to explore the intrinsic role of long-ranged parts of the interaction and their interplay with wavefunction quantum geometries in determining the ground states and the quantum phase transitions. For instance at the $\mC=1$ case, a transition from the Laughlin state to Wigner crystal is proposed to occur in the heterostructure of Dirac material/ type-II superconductor when tuning the range of Coulomb interaction~\cite{Liang_DiracNonuniformB}. Richer phase diagrams and phase transitions are expected to occur in high Chern bands and in multilayer \mr materials.

Our theory also opens new directions to the theory of FCI and FQH. Understanding the density operator and their correlations in flatband systems has been a long term topic~\cite{Qi_wanniermapping11,PhysRevB.90.165139,Jackson:2015aa,PARAMESWARAN2013816,Bernevig_Dalgebra12,Sheng:2011tr,Cecil_singlemodeapp,Murthy_FCI12}. As GMP algebra is tightly related to the neutral excitation~\cite{gmpb,gmpl,Haldanegeometry}, it is interesting to ask how Berry curvature fluctuation influences it and the possibility of nematic transitions tuned by geometry, which can be examined based on the theory presented here. Moreover, since in quantum Hall physics the GMP algebra is closely related to other responses such as Hall viscosity~\cite{Read_viscosity09,Read_viscosity11,Haldanegeometry,Grovmov_Son_PRX17,Bradlyn_Kubo12}, we believe ideal flatband is a nature place to explore this geometric response and beyond in flatband systems~\cite{Taylor_viscosity15,sinsei_viscosity15,Barry_latticeviscosity20}.

Furthermore, our work bridges condensed matter physics and mathematics. Many of our results are motivated and supported by intuitions and rigorous statements from both fields, in particular the quantum Hall and flatband physics, and the K\"ahler geometry in mathematics. We expect more insights from bridging these two fields in the future. Concretely, the Bergman kernel might be useful to reformulate the density algebra derived in this work and offers new perspectives to the emergent pseudopotential projectors. The coherent states in the emergent Hilbert space $\tmH$ and geometric quantization also deserve further exploration. Moreover, higher dimensional generalization of ideal flatbands is also an interesting future direction~\cite{HuZhang01,KARABALI02,Jie_FQHCP2}.

\begin{acknowledgements}
We are grateful to Jennifer Cano, Andrew J. Millis and Bo Yang for collaboration on $\mC=1$ ideal flatbands~\cite{JieWang_exactlldescription}. The work of S.K. was partly supported by the IdEx program and the USIAS Fellowship of the University of Strasbourg. Z.L. is supported by the National Key Research and Development Program of China through Grant No. 2020YFA0309200. The Flatiron Institute is a division of the Simons Foundation. This work is also supported by Center for Mathematical Sciences and Applications at Harvard University.

\emph{Note added}: during the preparation of the draft we are aware of the work by Junkai Dong {\it et al.}~\cite{junkaidonghighC22} which overlapped with part of our results.
\end{acknowledgements}

\appendix
\section*{--- Appendix ---}
\section{Some useful derivation details}
\subsection{Fourier coefficients}
In this section we discuss the derivation details for Eq.~(\ref{sym_omega1}) and Eq.~(\ref{sym_omega2}). Eq.~(\ref{sym_omega1}) follows directly from the following identity:
\begin{equation}
    \mathcal A_{\bm\sigma,\bm\sigma'}(\bm r) = \mathcal A_{\bm 0,\delta\bm\sigma}(\bm r+\bm a_{\bm\sigma})\times e^{\frac{i\mC}{2}\bm q_{\bm\sigma}\times\bm q_{\bm\sigma'} - \frac{i\mC}{2}\bm q_{\delta\bm\sigma}\cdot\bm a_{\bm\sigma}}.
\end{equation}
Therefore the Fourier coefficients $\omega_{\tbm b}(\bm\sigma,\bm\sigma')$ can all be reduced to $\omega_{\tbm b}(\bm 0,\delta\bm\sigma)$ which we simply denote as $\omega_{\tbm b}(\delta\bm\sigma)$ where $\delta\bm\sigma=\bm\sigma'-\bm\sigma$. Eq.~(\ref{sym_omega2}) follows from:
\begin{equation}
    \omega_{\tbm b}(\bm\sigma+\mC_ie_i) = -e^{\frac{i}{2\mC}\epsilon_{ij}\tbm b_j\times\bm q_{\bm\sigma}}\times\omega_{\tbm b+\epsilon_{ij}\tbm b_j}(\bm\sigma),\label{append_omega}
\end{equation}
which can be obtained by using the quasi-periodicity of $\mathcal B(\bm r)$ given in Eq.~(\ref{B_quasiperiodic}):
\begin{equation}
    \mathcal A_{\bm 0,\bm\sigma+\mC_ie_i} (\bm r) = -e^{-\frac{i}{2\mC}\tbm a_i\times(\bm r+\bm a_{\bm\sigma})}\times\mathcal A_{\bm 0,\bm\sigma}(\bm r).
\end{equation}

\subsection{Form factor}
In this section, we show the calculation details of the form factor $\mF^{\bm k\bm k'}_{\bm b}$ defined in the main text. For the $\bm b = \bm 0$ component, we simply denote as $\mF^{\bm k\bm k'}$. Using the general form of the single-particle wavefunction, we arrive at:
\begin{eqnarray}
    \mF^{\bm k\bm k'}_{\bm b} &=& \sum_{\bm\sigma\bm\sigma'}\int d^2\bm r~e^{-\frac{i\mC}{2}\bm q_{\bm\sigma}\times\bm q_{\bm\sigma'}}e^{-i\bm b\cdot\bm r}e^{\frac{i}{2}\left(\bm q_{\bm\sigma} - \bm q_{\bm\sigma'}\right)\cdot\bm r}\nonumber\\
    &\times& \mathcal A_{\bm\sigma\bm\sigma'}(\bm r)v^{\bm\sigma*}_{\bm k}(\bm r)v^{\bm\sigma'}_{\bm k'}(\bm r),
\end{eqnarray}
where $\mathcal A_{\bm\sigma\bm\sigma'}$ is defined in the main text. Then by using,
\begin{equation}
    v^{\bm\sigma}_{\bm k}(\bm r) = e^{-i\left(\bm k+\bm q_{\bm\sigma}/2\right)\cdot\bm r}e^{-\frac{i\mC}{2}\bm q_{\bm\sigma}\times\bm k}\Phi_{\bm k+\bm q_{\bm\sigma}}(\bm r),
\end{equation}
we arrive at the follows:
\begin{eqnarray}
    \mF^{\bm k\bm k'}_{\bm b} &=& \sum_{\bm\sigma\bm\sigma'}\sum_{\tbm b}\omega_{\tbm b}(\bm\sigma,\bm\sigma')e^{\frac{i\mC}{2}\left(-\bm q_{\bm\sigma}\times\bm q_{\bm\sigma'} + \bm q_{\bm\sigma}\times\bm k - \bm q_{\bm\sigma'}\times\bm k'\right)}\nonumber\\
    &\times& \langle\Phi_{\bm k+\bm q_{\bm\sigma}}| e^{\left(\tbm b-\bm b+\bm k-\bm k'+\bm q_{\bm\sigma}-\bm q_{\bm\sigma'}\right)\cdot\hat{\bm r}} |\Phi_{\bm k'+\bm q_{\bm\sigma'}}\rangle.
\end{eqnarray}
The second line is nothing but the LLL form factor. We summarize the expression for the form factor in below:
\begin{equation}
    \mF^{\bm k\bm k'}_{\bm b} = \sum_{\bm\sigma\bm\sigma'}\sum_{\tbm b}\omega_{\tbm b}(\bm\sigma,\bm\sigma')g^{\bm k\bm k'}_{\bm b-\tbm b}(\bm\sigma,\bm\sigma'),\label{append_mF}
\end{equation}
where functions $g$ and $f$ are:
\begin{eqnarray}
    g^{\bm k\bm k'}_{\tbm b}(\bm\sigma,\bm\sigma') &\equiv& e^{\frac{i\mC}{2}\left(-\bm q_{\bm\sigma}\times\bm q_{\bm\sigma'} + \bm q_{\bm\sigma}\times\bm k - \bm q_{\bm\sigma'}\times\bm k'\right)}f^{\bm k+\bm q_{\bm\sigma},\bm k'+\bm q_{\bm\sigma'}}_{\tbm b},\nonumber\\
    f^{\bm k\bm k'}_{\tbm b} &\equiv& \eta_{\tbm b}e^{\frac{i\mC}{2}(\bm k+\bm k')\times\tbm b}e^{\frac{i\mC}{2}\bm k\times\bm k'}e^{-\frac{\mC}{4}|\bm k-\bm k'-\tbm b|^2}.
\end{eqnarray}
We notice,
\begin{equation}
    g^{\bm k\bm k'}_{-\tbm b}(\bm\sigma,\bm\sigma') = \eta_{\tbm b}e^{-\frac{i\mC}{2}\bm k\times\tbm b}e^{-i\mC\bm q_{\bm\sigma}\times\tbm b} \times g^{\bm k+\tbm b,\bm k'}(\bm\sigma,\bm\sigma'),\label{append_rho01}
\end{equation}
where the index $\bm b$ is omitted for $g^{\bm k\bm k'}_{\bm b = \bm 0}$. It is easy to check that $g^{\bm k\bm k'}(\bm\sigma, \bm\sigma')$ depends only on the color difference:
\begin{equation}
    g^{\bm k\bm k'}(\delta\bm\sigma) = e^{-\frac{i\mC}{2}\bm q_{\delta\bm\sigma}\times(\bm k+\bm k')}e^{\frac{i\mC}{2}\bm k\times\bm k'}e^{-\frac{\mC}{4}|\bm k-\bm k'-\bm q_{\delta\bm\sigma}|^2}.\label{append_rho02}
\end{equation}

\subsection{Density operator}
In this section we derive the expression for the flatband density operator $\hat\rho_{\bm q} \equiv \sum_{\bm k}\mF^{\bm k+\bm q,\bm k}c^\dag_{\bm k+\bm q}c_{\bm k}$ and derive their algebra. First, by using Eq.~(\ref{sym_omega1}), Eq.~(\ref{append_mF}), Eq.~(\ref{append_rho01}) and Eq.~(\ref{append_rho02}), the explicit form for the density operator is:
\begin{equation}
    \hat\rho_{\bm q} = \sum_{\delta\bm\sigma}\sum_{\bm k,\tbm b}\eta_{\tbm b}e^{-\frac{i\mC}{2}(\bm k+\bm q)\times\tbm b}\omega_{\tbm b}(\delta\bm\sigma)g^{\bm k+\bm q+\tbm b,\bm k'}(\delta\bm\sigma)\cdot c^\dag_{\bm k+\bm q}c_{\bm k},\label{append_rho1}
\end{equation}
where $\sum_{\delta\bm\sigma}$ sums color from $\delta\sigma_i\in[0,\mC_i-1]$. Now we further simply the density operator from the $\tbm b-$lattice to the $\bm b-$lattice. By using Eq.~(\ref{sym_omega2}):
\begin{equation}
    \omega_{\tbm b}(\delta\bm\sigma) = \eta_{\tbm b'}e^{-\frac{i\mC}{2}\tbm b'\times\bm q_{\delta\bm\sigma}}\omega_{\bm b}[\bm q_{\delta\bm\sigma}-\tbm b'],
\end{equation}
where $\tbm b = \bm b + \tbm b'$, we arrive at the following steps of derivation:
\begin{widetext}
    \begin{eqnarray}
        \hat\rho_{\bm q} &=& \sum_{\delta\bm\sigma}\sum_{\bm k,\bm b,\tbm b'}\eta_{\bm b+\tbm b'}e^{-\frac{i\mC}{2}(\bm k+\bm q)\times(\bm b+\tbm b')}\times \eta_{\tbm b'} e^{-\frac{i\mC}{2}\tbm b'\times\bm q_{\delta\bm\sigma}}\omega_{\bm b}[\bm q_{\delta\bm\sigma}-\tbm b'] \times g^{\bm k+\bm q+\bm b+\tbm b',\bm k}[\bm q_{\delta\bm\sigma}]\cdot c^\dag_{\bm k+\bm q}c_{\bm k}\nonumber\\
        &=& \sum_{\delta\bm\sigma}\sum_{\bm k,\bm b,\tbm b'}\eta_{\bm b}e^{-\frac{i\mC}{2}(\bm k+\bm q)\times\bm b}\times\omega_{\bm b}[\bm q_{\delta\bm\sigma}-\tbm b']\times g^{\bm k+\bm q+\bm b,\bm k}[\bm q_{\delta\bm\sigma} - \tbm b']\cdot c^{\dag}_{\bm k+\bm q}c_{\bm k},
    \end{eqnarray}
\end{widetext}
where $g^{\bm k+\tbm b,\bm k'}(\bm q_{\delta\bm\sigma}) = e^{-\frac{i\mC}{2}\tbm b\times(\bm k-\bm q_{\delta\bm\sigma})} g^{\bm k,\bm k'}(\bm q_{\delta\bm\sigma}-\tbm b)$ and $\eta_{\tbm b}\eta_{\bm b}=\eta_{\tbm b+\bm b}e^{\frac{i\mC}{2}\tbm b\times\bm b}$ are used. We defined $g^{\bm k\bm k'}[\bm q_{\bm\sigma}] \equiv g^{\bm k\bm k'}(\bm\sigma)$. This simplifies the density operator into the following form:
\begin{equation}
    \hat\rho_{\bm q} = \sum'_{\delta\bm\sigma}\sum_{\bm k,\bm b}\eta_{\bm b}e^{-\frac{i\mC}{2}(\bm k+\bm q)\times\bm b}\omega_{\bm b}[\bm q_{\delta\bm\sigma}]~g^{\bm k+\bm q+\bm b,\bm k}(\delta\bm\sigma) c^\dag_{\bm k+\bm q}c_{\bm k},\label{append_rho2}
\end{equation}
where $\sum'_{\delta\bm\sigma}$ sums color over $\delta\sigma_i\in[0,\mC-1]$. Then using the boundary condition from Eq.~(\ref{def_kbc_uk}) $c^\dag_{\bm k+\bm q} = \eta_{\bm b}e^{-\frac{i\mC}{2}\bm b\times(\bm k+\bm q)}\cdot c^\dag_{\bm k+\bm q+\bm b}$, we arrive at the final result:
\begin{equation}
    \hat\rho_{\bm q} = \sum'_{\delta\bm\sigma}\sum_{\bm k,\bm b}\tilde\omega_{\bm b}[\bm q_{\delta\bm\sigma}]~g^{\bm k+\bm q+\bm b,\bm k}(\delta\bm\sigma)\cdot c^\dag_{\bm k+\bm q+\bm b}c_{\bm k}.\label{append_rho3}
\end{equation}
This can be rewritten as,
\begin{equation}
    \hat\rho_{\bm q} = \sum'_{\delta\bm\sigma}\sum_{\bm b}\tilde\omega_{\bm b}[\bm q_{\delta\bm\sigma}]\cdot\hat\rho^{\rm GMP}_{\bm q+\bm b}(\delta\bm\sigma),
\end{equation}
where
\begin{equation}
    \hat\rho^{\rm GMP}_{\bm q}(\delta\bm\sigma) \equiv \sum_{\bm k} g^{\bm k+\bm q,\bm k}(\delta\bm\sigma)\cdot c^\dag_{\bm k+\bm q}c_{\bm k}.
\end{equation}

We now prove $\hat\rho^{\rm GMP}_{\bm q}(\delta\bm\sigma)$ satisfies a closed algebra. To start, we note that independent on particle statistics, the commutator of the density operator is
\begin{eqnarray}
    && [\hat\rho^{\rm GMP}_{\bm q_1}(\delta\bm\sigma_1), \hat\rho^{\rm GMP}_{\bm q_2}(\delta\bm\sigma_2)] = \sum_{\bm k} c^\dag_{\bm k+\bm q_1+\bm q_2}c_{\bm k}\label{append_densityalgebra1}\\
    &\times& \left[g^{\bm k+\bm q_1+\bm q_2,\bm k+\bm q_2}(\delta\bm\sigma_1)g^{\bm k+\bm q_2,\bm k}(\delta\bm\sigma_2) - (1\leftrightarrow2)\right].\nonumber
\end{eqnarray}
It is easy to verify that $g$ satisfies a ``chain rule'':
\begin{eqnarray}
    & & \frac{g^{\bm k+\bm q_1+\bm q_2,\bm k+\bm q_2}(\delta\bm\sigma_1)\cdot g^{\bm k+\bm q_2,\bm k}(\delta\bm\sigma_2)}{g^{\bm k+\bm q_1+\bm q_2,\bm k}(\delta\bm\sigma_1+\delta\bm\sigma_2)},\nonumber\\
    &=& e^{-\frac{i\mC}{2}\bm q_{\delta\bm\sigma_1}\times\bm q_{\delta\bm\sigma_2}} e^{\mC\left(q_1-q_{\delta\bm\sigma_1}\right)^*\left(q_2-q_{\delta\bm\sigma_2}\right)}.
\end{eqnarray}
We see the right hand side is $\bm k$-independent: this is nontrivial and leads to the closed density algebra shown in the main text after plugging into Eq.~(\ref{append_densityalgebra1}).

\bibliography{ref.bib}

\end{document}